\documentclass[useAMS,usenatbib]{mn2e}
\usepackage{txfonts,graphicx} 
\usepackage{natbib,longtable,url,supertabular} 
\usepackage[usenames]{color}

\newcommand\ion[2]{#1$\;${\scshape{#2}}} 
\newcommand{\pone}{{\mathcal{P}}_{\rm LTE,Fe}}
\newcommand{\ptwo}{{\mathcal{P}}_{\rm NLTE,opt}}
\newcommand{\tone}{T_{\rm LTE,Fe}}
\newcommand{\ttwo}{T_{\rm NLTE,opt}}
\newcommand{\gone}{\log g_{\rm LTE,Fe}}
\newcommand{\gtwo}{\log g_{\rm NLTE,opt}}

\newcommand{\Teff}{\ensuremath{T_{\mathrm{eff}}}}     
\newcommand{\logg}{\ensuremath{\log g}}               

\newcommand{\logeFesun}[1]{\log\varepsilon_{\rm Fe, \odot}}
\newcommand{\logeFe}[1]{\log\varepsilon_{\rm Fe, \star}}
\newcommand{\logemean}[1]{\log\varepsilon_{\rm mean}}

\newcommand{\apj}{ApJ}                    
\newcommand{\apjs}{ApJS}
\newcommand{\apjl}{ApJL}
\newcommand{\aap}{A{\&}A}

\newcommand{\mnras}{MNRAS}
\newcommand{\aj}{AJ}
\newcommand{\araa}{ARAA}

\newcommand{\ssr}{Space~Sci.~Rev.}

\title[Ages, masses,  and distances of cool stars]{Bayesian  analysis of ages,
  masses, and  distances to cool stars with  non-LTE spectroscopic parameters}
\author[A. Serenelli]{Aldo M. Serenelli$^{1,2}$\thanks{Email:aldos@ice.csic.es},
Maria  Bergemann$^2$,   Gregory   Ruchti$^{2,3}$, Luca Casagrande$^4$ \\  
$^1$ Institute of Space Sciences  (IEEC-CSIC), Campus UAB, Fac. Ci\`encies,
Torre
  C5  parell 2,  08193, Bellaterra,  Spain \\  $^2$ Max  Planck  Institute for
  Astrophysics, Karl-Schwarzschild Str. 1, 85741, Garching, Germany 
\\ $^3$ Lund Observatory, Department of Astronomy and Theoretical Physics, Box
43, SE-221 00 Lund, Sweden \\ $^4$
  Research School of Astronomy \& Astrophysics, Mount Stromlo Observatory, The
Australian 
  National University, ACT 2611, Australia}

\begin{document}

\date{Accepted Date. Received Date; in original Date}

\pagerange{\pageref{firstpage}--\pageref{lastpage}} \pubyear{2012}

\maketitle

\label{firstpage}

\begin{abstract}
For studies of  Galactic evolution, the accurate characterization  of stars in
terms of their evolutionary stage  and population membership is of fundamental
importance.  A  standard approach relies  on extracting this  information from
stellar  evolution  models but  requires  the  effective temperature,  surface
gravity, and metallicity of a star obtained by independent means.  In previous
work, we  determined accurate effective  temperatures and non-LTE  $\logg$ and
[Fe/H] (NLTE-Opt)  for a large  sample of metal-poor  stars, $-3 <$  [Fe/H] $<
-0.5$, selected  from the  RAVE survey.   As a continuation  of that  work, we
derive  here their masses,  ages, and  distances using  a Bayesian  scheme and
GARSTEC  stellar  tracks.  For  comparison,  we  also  use stellar  parameters
determined  from the widely-used  1D LTE  excitation-ionization balance  of Fe
(LTE-Fe).   We find  that the  latter leads  to  systematically underestimated
stellar ages, by 10-30\%,  but overestimated masses and distances.  Metal-poor
giants  suffer  from  the   largest  fractional  distance  biases  of  $70\%$.
Furthermore,  we  compare  our  results   with  those  released  by  the  RAVE
collaboration   for   the  stars   in   common  (DR3,   \citealp{zwitter:2010,
  seibert:2011}).   This  reveals $-400$  to  $+400$  K  offsets in  effective
temperature, $-0.5$ to $1$ dex offsets  in surface gravity, and $10$ to $70\%$
in distances. The systematic trends  strongly resemble the correlation we find
between the NLTE-Opt and LTE-Fe  parameters, indicating that the RAVE DR3 data
may be affected  by the physical limitations of the  1D LTE synthetic spectra.
Our  results bear on  any study,  where spectrophotometric  distances underlie
stellar  kinematics.   In particular,  they  shed  new  light on  the  debated
controversy  about   the  Galactic  halo  origin  raised   by  the  SDSS/SEGUE
observations.
\end{abstract}
%
%
\section{Introduction}
One  of the major  problems in  modern astrophysics  is to  determine physical
parameters of stars,  masses and ages, and their  kinematics.  Whereas various
methods exist, most of them are  not suitable for large stellar samples.  Thus
the  most  common approach  is  fitting  to  stellar evolution  models,  i.e.,
comparison of an observed location of a  star in the $\Teff - M_V$ plane
with  isochrones   or  evolutionary  tracks   for  a  given   metallicity  and
$\alpha$-enhancement.  This  method demands stellar  atmospheric parameters as
an      input,     which      can     be      derived      from     photometry
\citep[e.g.,][]{holmberg:2007,casagrande:2011}         or         spectroscopy
\citep{valenti:2005}, or  some combination thereof  \citep[e.g. application to
  SDSS/SEGUE  data in][]{lee:2008}.  The  drawback of  photometry  is that  it
depends on reddening, which is  difficult to estimate, and demands calibration
of  metallicity and  alpha-enhancement on  spectroscopic  data.  Additionally,
information  on  gravity  is  not  obtained directly,  a  quantity  needed  to
accurately determine  distances.  The alternative and preferable  method is to
use  spectroscopy directly,  that  allows a  consistent  determination of  all
atmospheric quantitites,  effective temperature, surface  gravity, and element
abundances, simultaneously.  Thus, it minimizes  the the total error caused by
various sources  of uncertainties inherent  to different methods  of parameter
determination.

With   the   advent   of    large-scale   stellar   surveys   such   as   RAVE
\citep{steinmetz:2006},       GCS       \citep{nordstrom:2004},      SDSS/SEGUE
\citep{yanny:2009},  GAIA-ESO  \citep{gilmore:2012}  much attention  has  been
devoted to  developing sophisticated statistical  methods to best  exploit the
available spectroscopic  and photometric  information. The basic  technique of
isochrone fitting  (see e.g. \citealp{ng:1998,liu:2000}), or  a combination of
complementary  techniques   \citep{lachaume:1999},  have  evolved   into  more
systematic  methods  that  can be  applied  to  very  large amounts  of  data.
Different    alternatives   based    on   a    frequentist    approach   (e.g.
\citealt{breddels:2010})   or    Bayesian   analysis   of    ever   increasing
sophistication                               \citep{jorgensen:2005,takeda:2007,
  casagrande:2011,burnett:2010} have  been presented and  their advantages and
disadvantages discussed \citep{pont:2004,jorgensen:2005}.

As a  result, spectrophotometric determination of masses,  ages, and distances
for  late-type stars  is  nowadays routinely  applied  in large-scale  stellar
surveys  and, although  with different  implementations, lies  at the  core of
almost all  studies aimed at  characterizing stellar populations in  the Milky
Way,  both  in  the  intrinsic  stellar  properties  such  as  age-metallicity
relations   \citep{holmberg:2007,casagrande:2011}   as   well  as   in   their
kinematical    properties    \citep{carollo:2010,schonrich:2011,burnett:2011}.
However,  in these  studies  little or  no  attention has  been  given to  the
physical  accuracy of  stellar atmospheric  parameters, which,  in one  way or
another,  always rely  on  models describing  radiative  transport in  stellar
atmospheres.   In spectroscopy,  all  theoretical grids  of synthetic  stellar
spectra for  late-type stars  have been computed  based on the  assumptions of
local thermodynamic and 1D  hydrostatic equilibrium (hereafter 1D LTE).  These
are,   for    example,   the   widely-used   synthetic    spectra   grids   by
\citet[][]{munari:2005}, which rely on the  Kurucz 1D LTE model atmosphere and
radiative  transfer  codes.   However,  recent  observational  and  theoretial
studies revealed substantial systematic  biases in basic stellar parameters at
low  metallicity  and/or  gravity  caused  by  the breakdown  of  LTE  and  1D
hydrostatic     equilibrium    approximations    \citep[e.g.,][]{asplund:2005,
  bergemann:2012}.  It must be pointed out that these biases are
ubiquitous, and affect the results obtained from the analysis of high- and
low-resolution spectra. The reason is that departures from 1D HE and LTE often
change equivalent  widths (or, equivalently, line indices), and  consequently 
abundances inferred from  a diagnostic spectral feature. This  effect does not
vanish with a decreasing spectral resolution, which merely alters the
line  profile shape without affecting the total energy absorbed in a line.
Although no detailed quantitative estimates have been performed so far, one may
also expect a non-negligible impact on certain photometric indices. As a
consequence, it is completely unknown whether, and how, systematic errors  in
basic stellar parameters caused  by the assumptions of 1D  LTE impact the 
inferred stellar properties,  that is, in our context, ages, masses, and
distances.

In  \citet[][hereafter  Paper 1]{ruchti:2012},  we  presented a  comprehensive
study  of systematic  errors  in the  determination  of spectroscopic  stellar
parameters arising  from using  the classical method  of excitation-ionization
balance  of  \ion{Fe}{I}  and  \ion{Fe}{II}   lines  based  on  1D  LTE  model
atmospheres.  Based on the results for the large sample of metal-poor stars
($-3 <$ [Fe/H] $< -0.5$) selected  from the  RAVE  catalogue, we  concluded 
that the  impact of  using inaccurate effective temperature scale and neglecting
the effects of non-local thermodynamic  equilibrium  (NLTE)  on  the 
determination  of  gravities  and metallicities  leads  to systematic  biases 
of up  to  $400$  K in  effective temperature, $1.5$  dex in  surface gravity
and  $0.5$ dex in  metallicity\footnote{While the methods developed in
the Paper 1 are best suited for application to high-resolution spectra, they can
be, in principle, extended to lower-resolution data if the wavelength coverage
is sufficient to measure the H lines or accurate $\Teff$ can be obtained by
independent means, i.e., from high-quality photometry with a careful correction
for reddening. A more reasonable way is to resort to standard methods
relying on global $\chi^2$ minimization, as e.g. in \citet{zwitter:2008},
using, however,the synthetic grids computed completely in NLTE. Such a code is
currently being developed by one of the co-authors and will be applied to a
large dataset of low-resolution spectra.}. In this work, we use results from
Paper 1 to analyze the impact of using physically accurate  spectroscopic
stellar parameters on  the determination of their masses, ages, and distances.

Our paper is structured as follows: Section~\ref{sec:obs} gives details on the
observational data and  the different techniques used in  the determination of
basic   spectroscopic  parameters.    Section~\ref{sec:models}   is  a   brief
description of the grid of stellar evolution models computed for this work. In
Section~\ref{sec:bayes}  we describe  the novel  Bayesian method  we  apply to
obtain masses, ages and,  distances, and discuss the limitations, particularly
imposed by  the degeneracy  of stellar models  on the observational  plane. We
present the  results in Section~\ref{sec:results} and compare  them with those
published by the RAVE collaboration.  Finally, in Section~\ref{sec:summary} we
summarize  the impact  of  our findings  in  the broader  context of  Galactic
structure studies  and emphasize the  absolute need of  accurate spectroscopic
stellar parameters in  order to fully exploit the  enormous amount and quality
of data emerging from large-scale stellar surveys.

%
%
\section{Observed      stars,      model      atmospheres     and      stellar
  parameters}{\label{sec:obs}} 

The observed dataset  was taken from \citep{ruchti:2011} and  it comprises the
core of the  sample used in \citet[][hereafter Paper  1]{ruchti:2012} as well.
These stars were originally selected for high-resolution observations based on
data   obtained   by   the    Radial   Velocity   Experiment   Survey   (RAVE;
\citealp{steinmetz:2006}) in order  to study the metal-poor thick  disk of the
Milky Way.  High  signal-to-noise spectra for these stars  were obtained using
high-resolution echelle spectrographs ($R \geq 30\,000$ and S/N $\sim 100$ per
pixel)  at several facilities  around the  world.  For  the present  work, the
final sample contains 254 stars in the metallicity range \hbox{$-3.0 <$ [Fe/H]
  $<  -0.5$}. With respect  to the  full sample,  we have  excluded horizontal
branch candidates because the grid of stellar models used here extends only up
to the tip  of the red giant  branch (RGB).  Also, stars with  \hbox{$\logg <
  0.5$} have not  been considered to avoid large  extrapolations from the grid
of NLTE corrections that extends down to $\logg =1$
\citep{lind:2012}\footnote{The NLTE abundance corrections for the Fe I
lines are available for the full grid of \citet{lind:2012} under
\url{www.inspect-stars.net}.} Stellar  parameters  were  determined   using
two  different  techniques,  as described in detail in Paper 1 and summarized
below.

\begin{figure*}
\begin{center}
\includegraphics[scale=.40]{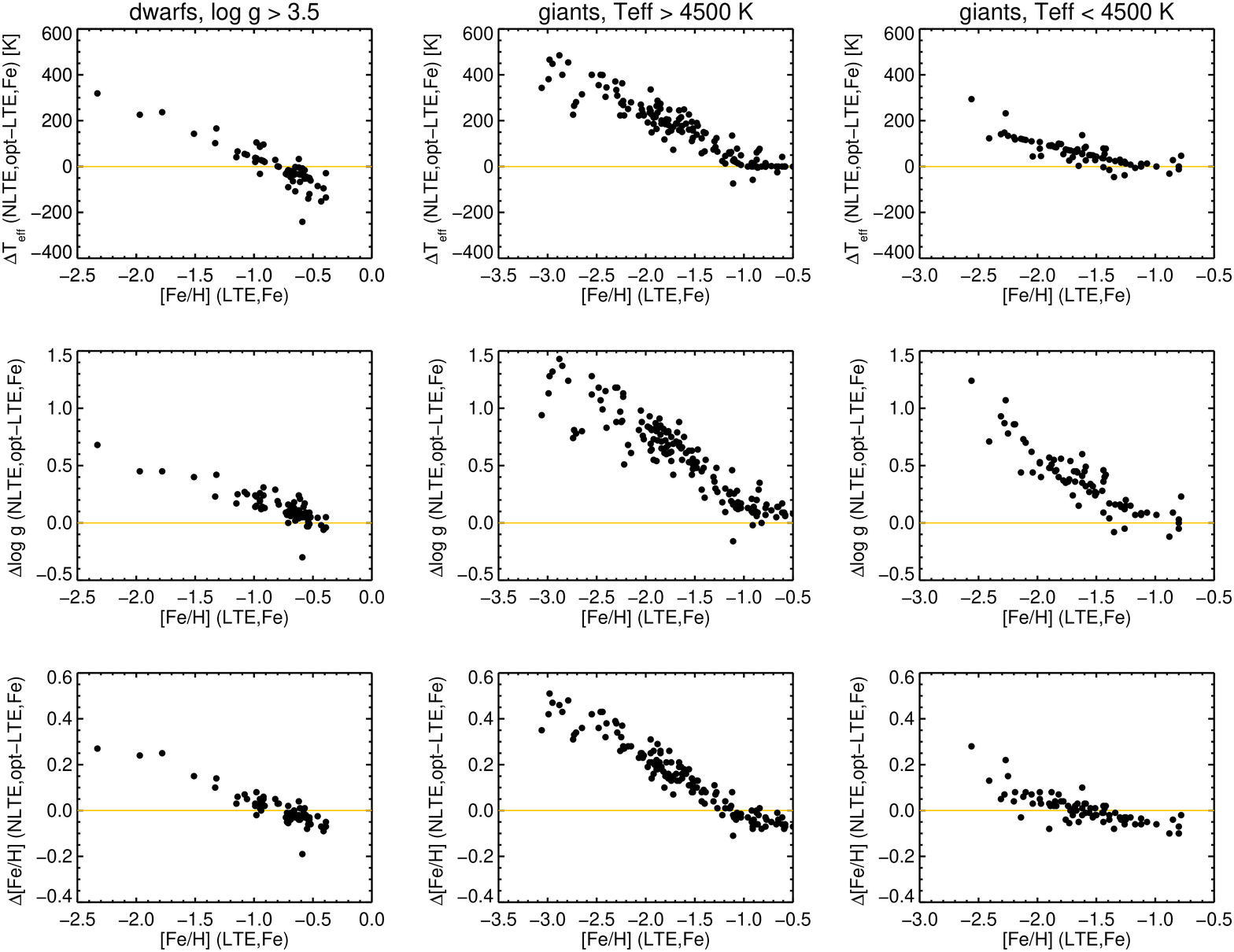}
\caption{Comparison of stellar parameters  derived using the LTE-Fe method and
  the  NLTE-Opt  stellar parameters  for  our  sample of  metal-poor
  stars. Differences in temperature,  gravity, and metallicity show a large
  systematic increase with decreasing metallicity. See Paper 1 for full details.
\label{fig:nltepar}}
\end{center}
\end{figure*}

First,  we  apply  the   classical  method  of  1D  LTE  excitation-ionization
equilibrium of \ion{Fe}{i} and  \ion{Fe}{ii}, which relies on the minimization
of the  abundance trends with reduced  line equivalent widths  and lower level
excitation potential.  $\Teff$ and  microturbulence are optimized to establish
excitation balance, and $\logg $ and [Fe/H] are found by simultaneously
requiring ionization balance. Notwithstanding  the physical limitations due to
the assumptions of 1D LTE, the method is still in widespread use today because
of its simplicity and easiness of implementation.

In the second approach, substantial  efforts were made to improve the
accuracy of basic  stellar parameters  in the attempt  to minimize  the
above-mentioned systematic  errors.  Firstly,  effective  temperatures were 
determined  from different  methods  available  for  cool  stars (such  as  the 
Balmer  lines, semi-empirically calibrated photometry). The  final, optimal,
$\Teff$ is that derived  from the  weighted averaging of  different methods. In 
a second step, we recomputed gravities and metallicites relaxing the LTE
assumption and adopting the  new $\Teff$ scale. The  NLTE corrections were 
applied to \ion{Fe}  {i} abundances  using the  new wrapper  code to  MOOG.
The typical errors are  $\sigma(\Teff) \approx 60 -  100$ K, $\sigma(\logg)
\approx 0.1$ dex,  and  $\sigma(\rm{[Fe/H]}) \approx  0.1$~dex.  Note  that 
these are  the absolute  errors  (systematic +  internal)  on  the  accuracy of 
our  stellar parameters.   Similar   errors  bars   are   typically   quoted  in
  standard techniques.  However, usually  those  represent the  internal 
precision of  a method  only. The results  obtained  by the 
LTE-Fe method  are affected by much larger systematic errors. In the Paper 1, we
contrasted our results for stars with surface gravities and temperatures
determined by independent techniques, i.e., from accurate \textit{Hipparcos}
parallaxes and interferometry. We found that while the LTE-Fe surface gravities
were systematically too low by $-0.32 \pm 0.39$ dex, the NLTE-Opt surface
gravities were in agreement with astrometric values to $0.02 \pm 0.11$ dex. This
supports the accuracy of the new method.

The spectroscopic stellar parameters obtained using both methods described above
are  compared in Fig.~\ref{fig:nltepar}.  Hereafter, these  two sets of stellar
parameters are  referred to  as $\pone$  (LTE-Fe method)  and $\ptwo$ (NLTE-Opt
method). The  major conclusion to be drawn from  these plots is that the
classical  method of 1D~LTE excitation-ionization balance  of Fe leads to
systematic errors in all three basic stellar parameters:

\begin{itemize}

\item $\Teff$ (Fig.~\ref{fig:nltepar}, top panel): the LTE-Fe $\Teff$
values are too low by \hbox{100 to 400~K} for all stars below [Fe/H] $\sim -1$;
for dwarfs above this metallicity, the effective temperatures are usually
over-estimated.  Note that  the LTE-Fe  $\Teff$  values for  very cool  giants
appear to be reasonably accurate for [Fe/H] $> -1.5$.

\item $\logg$ (Fig.~\ref{fig:nltepar}, middle panel): surface gravities
are usually under-estimated. The largest effects is seen for warm giants, $\Teff
> 4500$ K, which are biased by up to $1.5$ dex at [Fe/H] $\sim -3$ and by $0.5$
dex at [Fe/H] $\sim -1.5$; for metal-poor dwarfs, gravities are under-estimated
by $\sim 0.4$ dex. For stars with [Fe/H] $\gtrsim -1$, the errors are within
$0.2$ dex.

\item [Fe/H] (Fig.~\ref{fig:nltepar}, bottom panel): the error is
nearly a linear function of metallicity itself, such that for dwarfs and giants
with [Fe/H] $\gtrsim -1$ the error in [Fe/H] is $\sim -0.1$dex. The error
changes sign and increases to $+0.2$ dex at [Fe/H] $\sim -1.5$, while at [Fe/H]
$\sim -3$ the LTE metallicity is too low by $0.5$ dex for giants and $0.3$ dex
for dwarfs.

\end{itemize}
Is it seen that the systematic effects are metallicity- and gravity-dependent,
which is reflected in a notable spread  of $\ttwo - \tone$ and  $\gtwo  - \gone$
at a given  $\logg$  and $\Teff$.

\begin{figure*}
\begin{center}
\hbox{
\includegraphics[width=85mm]{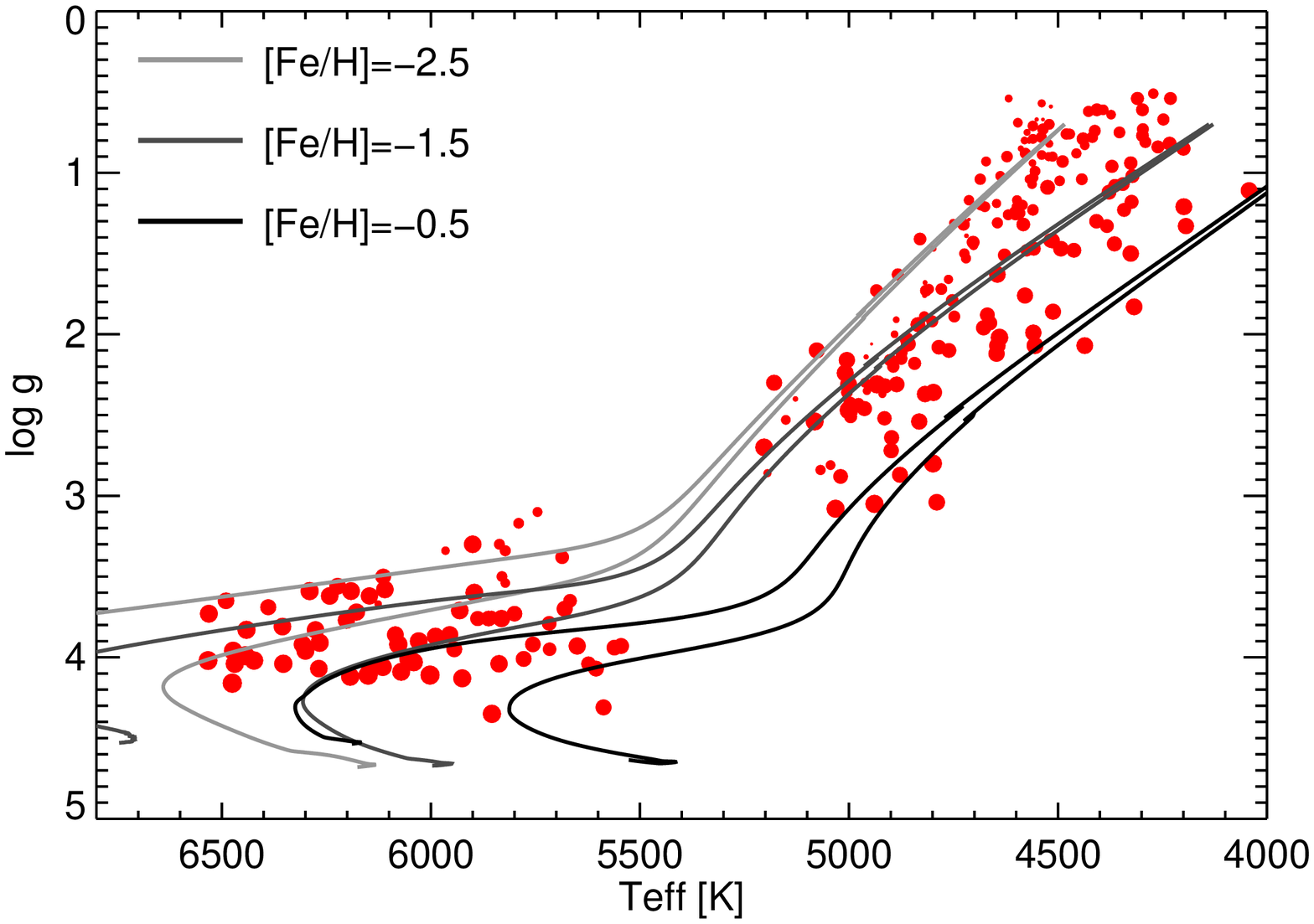}
\includegraphics[width=85mm]{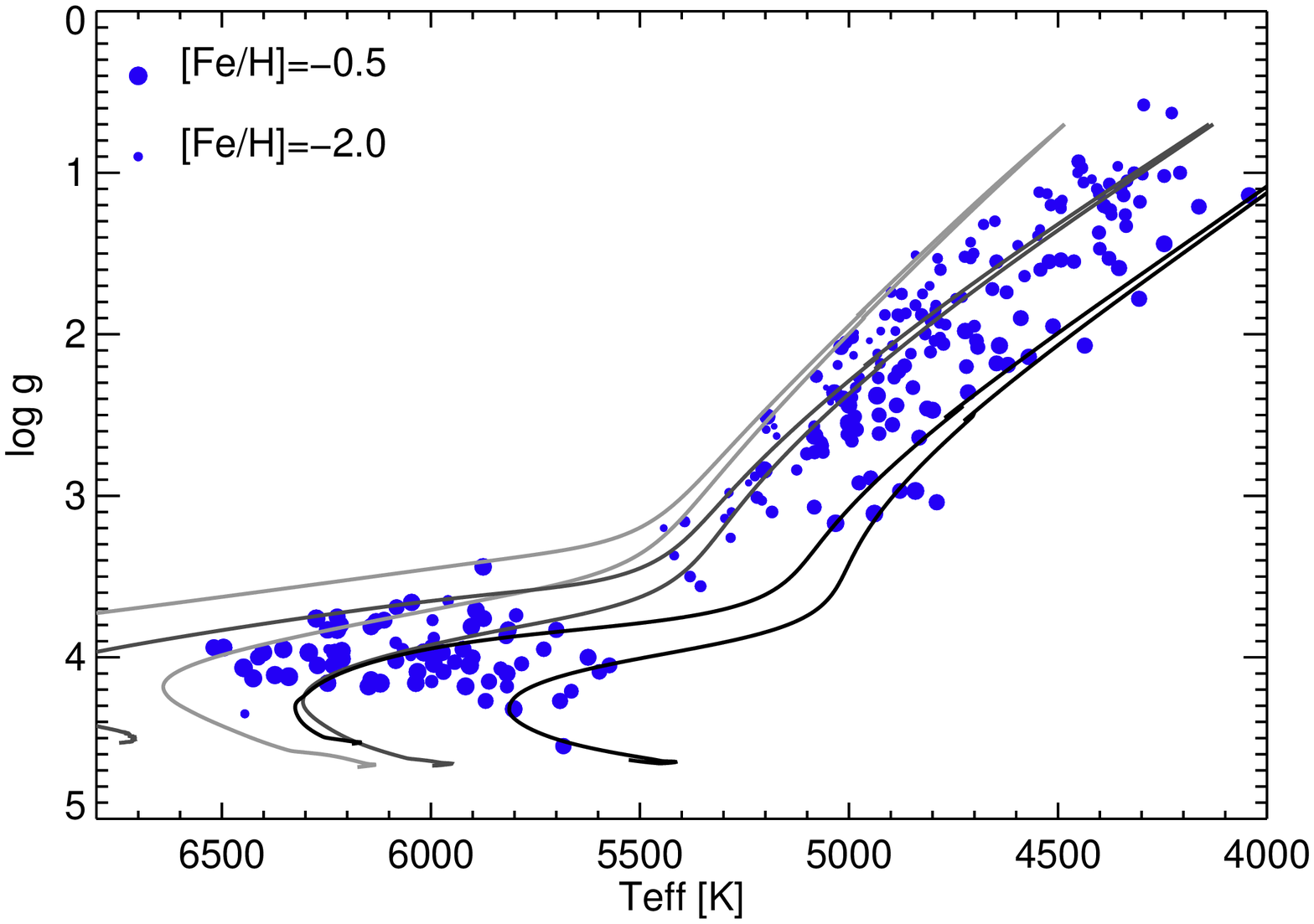}
}
\caption{The location of the stars for the parameter sets $\pone$ (left panel)
  and $\ptwo$  (right panel)  on the HRD. For comparison, the reference
stellar evolutionary  tracks of 0.8 and 1~M$_\odot$ and metallicities indicated
in the legend are overplotted. Symbol sizes are proportional to the
metallicity of  the observed stars  in both panels  as indicated in  the right
plot.
\label{fig:tracks_data}}
\end{center}
\end{figure*}

Fig.~\ref{fig:tracks_data}  compares  the  location   of  the  stars  for  the
parameter sets $\pone$ (left panel) and $\ptwo$ (right panel) on the $\Teff$ -
$\logg$  plane together  with reference  stellar evolutionary  tracks computed
using the \texttt{GARSTEC} code \citep[][see next section for a description of
  the grid  of stellar models]{weiss:2008}.  Tracks are identical in  the left
and right plots.  Symbol sizes are proportional to [Fe/H]  as indicated in the
right figure. Even though the tracks are shown only for visual guidance, it is
clear that  the $\ptwo$ parameters trace  the morphology of  the turn-off much
better.  Results for giants  are also  quite encouraging.  For each  star with
$\Teff < 5300$~K, we have searched  the entire grid of stellar models the find
the  best fit  to $\pone$  and $\ptwo$  parameters as  defined by  the minimum
$\chi^2$. In the case of  $\pone$ parameters, the average minimum $\chi^2$ for
these stars is 1.9, whereas if $\ptwo$ parameters are used it is 0.4. Clearly,
$\ptwo$ parameters lead to an improved overall agreement with stellar tracks.

%
%
\section{Evolutionary tracks}{\label{sec:models}}

Stellar  models   have  been  computed  with   \texttt{GARSTEC}  and  detailed
information  about  numerical  aspects  of  the code  and  the  input  physics
available  can be found  in \citet{weiss:2008};  here we  just provide  a very
brief  summary of  the most  relevant input  physics included  in  the stellar
models.

Regarding         microphysics,        we         have         used        the
FreeEOS\footnote{http://freeeos.sourceforge.net}     equation     of     state
\citep{cassisi:2003}  and  the  ${\rm ^{14}N(p,\gamma)^{15}O}$  reaction  rate
recommended by LUNA \citep {marta:2008}.  Convection is accounted for by using
the  standard  mixing length  theory  \citep{kippenhahn:1990}  and the  mixing
length parameter $\alpha_{\rm  MLT}$ is taken from the  calibration of a solar
model (see below).  Convective overshooting  is modeled as a diffusive process
following  the formulation  by (\citealt{freytag:1996}  and see  also  Eq.3 in
\citealt{weiss:2008}) and the free parameter  $f$ is fixed to 0.02. This value
represents a moderate  amount of overshooting on the  main sequence comparable
to the canonical  0.2~${\rm H_P}$\footnote{H$_P=P/g\rho$ is the pressure
scale  height,  where  $P$  and  $\rho$  are  the  pressure  and  gas  density
respectively.}.  
However, overshooting has to be limited in small convective 
cores\footnote{For example, a solar model  computed with the   canonical  amount
of  overshooting  of  0.2~${\rm H_P}$  still  has, at  the   present solar age, 
a convective core.  This is  ruled out from helioseismic   results.}.   In
\texttt{GARSTEC}  this  is now  achieved  by implementing  a geometric  cutoff
where  the extension  of the  overshooting  region decreases quadratically with
the  ratio between the size of the  convective core and the pressure scale
height at its boundary (see \citealt{magic:2010} for details).

In order to  fix the mixing length parameter and  the reference composition of
our   models  we   have  calibrated   a   standard  solar   model  using   the
\citet{grevesse:1998}  solar composition, from  which the  present-day surface
$(Z/X)_\odot=0.0229$  and $\log{\epsilon_{\rm  Fe}}=7.50$. The  latter defines
${\rm [Fe/H]= 0}$  in our models. From the  calibration we obtain $\alpha_{\rm
  MLT}=1.811$ and the initial solar helium abundance $Y_{\rm INI}=0.26896$ and
metallicity  $Z_{\rm   INI}=0.01876$.  Due  to  the   effects  of  microscopic
diffusion, the initial solar composition corresponds to ${\rm [Fe/H]= +0.06}$.

For the present work,  we have computed a dense grid of  stellar models in the
mass   range   $0.6   \leq   M/{\rm   M_\odot}   \leq   1.4$   in   steps   of
0.01~M$_\odot$\footnote{The  stellar  evolution models  can  be provided  upon
  request  to the  main  author}. The  initial  composition of  the models  is
computed assuming  a cosmic  helium-to-metal enrichment $\Delta  Y /  \Delta Z
=1.4$ anchored to  the initial solar composition. The  initial Fe abundance in
the  grid spans  the range  ${\rm -5.0  \leq [Fe/H]  \leq +0.5}$  in  steps of
0.1~dex  for   ${\rm  [Fe/H]  <  0.0}$   and  0.05~dex  for   ${\rm  [Fe/H]  >
  0.0}$ \footnote{The  initial [Fe/H] in  the evolutionary tracks  is 0.06~dex
  higher than  quoted on  the label.   We keep this  nomenclature in  order to
  associate [Fe/H]=  0 evolutionary tracks  to those with  initial composition
  equal to initial solar composition.}. The set of tracks on which we base the
results presented in the  following sections include microscopic diffusion and
a  0.4~dex  $\alpha$-element enhancement  for  ${\rm  [Fe/H]  \leq -0.6}$.   A
maximum depletion of metals due  to gravitational settling of about 0.4~dex is
obtained for  stars typical  of the turn-off  in metal  poor ([Fe/H]~$\sim-2$)
globular clusters.  No additional extra  mixing to those mentioned is included
in the models.  A detailed discussion on  this is of no relevance to this work
because  we are  interested in  assessing  the effect  of input  spectroscopic
parameters in  the determination of  stellar properties, and these  results do
not  depend on  the reference  stellar tracks  used. Models  are  evolved from
pre-main  sequence up  to  the tip  of  the RGB  or the  age  $\tau =30$  Gyr,
whichever occurs first.

%
%
\section{Bayesian analysis}\label{sec:bayes}

Bayesian inference offers a powerful  way to characterize a system by allowing
the  incorporation of  prior knowledge  into  the statistical  analysis in  an
objective  way.   In  the  context  of our  problem,  i.e.   determination  of
fundamental  stellar  parameters, Bayesian  methods  have  been discussed  and
developed  by various  authors.  Of  particular pedagogical  interest  are the
articles by  \citet{pont:2004} and \citet{jorgensen:2005} and  more recent and
sophisticated analysis with applications to large stellar surveys can be found
in  \citet{burnett:2010},  \citet{burnett:2011}  and  \citet{casagrande:2011}.
Our  implementation  of Bayesian  inference  of  basic  stellar parameters  is
described in the next section.

\subsection{The Method}{\label{sec:method}}

In Bayesian  analysis, the stellar parameters  we want to  determine, e.g mass
and age, are random variables  for which the probability distribution function
(PDF)  needs to  be  found. In  particular,  the posterior  PDF,  that is  the
probability of  a model being  correct given the  observed data and  the prior
knowledge, is expressed as
\begin{equation}
\begin{array}{lcl}
p(M,\tau,{\rm [Fe/H]}~|~T_{\rm eff, obs}, ...)  & \propto &
p (M,\tau,{\rm [Fe/H]})  \\
 & & \times \ L(T_{\rm eff, obs}, ...~|~M,\tau,{\rm [Fe/H]})
\end{array}
\label{eq:bayes}
\end{equation}
where the function $p (M,\tau,{\rm [Fe/H]})$ on the right-hand side represents
the  prior probability  of the  stellar  parameters and  $L(T_{\rm eff,  obs},
...~|~M,\tau,{\rm  [Fe/H]})$  the  likelihood,  i.e. the  probability  of  the
observed  parameters given  a  model  characterized by  a  set of  observables
D$_j(M, \tau, {\rm Fe/H]})$. Assuming  independence and Gaussian errors in the
  observed data, the likelihood is
\begin{equation}
L   =  \prod_j   \frac{1}{\sqrt(2\pi)  \sigma_j}   \  \times   \  \exp{\left(-
  \frac{\left(D_j^{\rm obs}  - D_j^{\rm mod}({\rm  M, \tau, [Fe/H]})\right)^2}
       {2 \sigma_j^2} \right)}.
\end{equation}
In our  case, $D$ consists of  three parameters: $\Teff$,  $\logg$, and
[Fe/H].   Other  parameters,  like  photometric information,  can  be  readily
implemented, but we do not consider them here.

The prior probability density of model parameters can be written as
\begin{equation}
p  (\tau,M,{\rm  [Fe/H]})  =  p(\tau)  \times  p({\rm  [Fe/H]}~|~\tau)  \times
p(M~|~{\rm [Fe/H]},\tau) 
\end{equation}
For the  prior on mass,  we assume a  standard Salpeter initial  mass function
(IMF) independent  of age  and metallicity. For  metallicity we assume  a flat
prior, i.e.   that at all  ages all metallicities  are {\em a  priori} equally
probable.  This  simple choice is  justified in this  work because we  are not
interested in  studying age-metallicity relations for  stellar populations and
because sufficiently reliable estimates of metallicity are available. Finally,
for the age we also assume a  flat prior, i.e.  that star formation rate (SFR)
is constant, but we set a cut at $\tau_{\rm max}=15$~Gyr. Therefore
\begin{equation}
p(M) \sim m^{-2.35},
\qquad m \in [0.6, 1.4] ~ M_{\odot},
\label{eq:imf}
\end{equation}
\begin{equation}
p(\tau) \sim 1/\tau_{\rm max},
\qquad \tau \in [0, 15] ~{\rm Gyr}.
\label{eq:sfr}
\end{equation}
More  complex  priors  can  be  elaborated,  but our  main  goal  here  is  to
investigate the  differences in masses, ages,  and distances due  to the input
spectroscopic parameters rather than to characterize stellar populations.

Finally,   the   proportionality  constant   in   Eq.~\ref{eq:bayes}  is   the
normalization   of  the   posterior  PDF   such  that   $\iiint  p(M,\tau,{\rm
  [Fe/H]}~|~T_{\rm eff, obs}, ...) \,{\rm dM\,d\tau\,d[Fe/H]}=1$, and it is of
no relevance for the estimation of parameters and uncertainties.

The PDF  of a stellar parameter $x$  that takes the values  $x_{\rm tr}$ along
evolutionary tracks is given by
\begin{equation}
\begin{array}{lcl}
p(x~|~T_{\rm eff, obs}, ...) & = & \iiint \delta\left(x - x_{\rm tr}\right) \\
& & p(M,\tau,{\rm [Fe/H]}\,|\,T_{\rm eff, obs}, ...)
\,dM\,d\tau\,d{\rm [Fe/H]}
\label{eq:ageprob}
\end{array}
\end{equation}
where the integration is over all evolutionary tracks. 

\subsection{General properties}

A key advantage of Bayesian inference  is that, as expressed before, it allows
using  prior  knowledge  about  the  problem  being  considered.   Priors  are
particularly important when a quantity of interest, affected by the prior, has
noticeable changes  within the range of observational  uncertainties where the
likelihood does not  change much.  A typical example,  originally discussed by
\citet{pont:2004}  and  further  expanded  by \citet{jorgensen:2005},  is  the
different evolutionary speed  of stellar models that populate  the same region
of observed parameters.  When this is  not accounted for, as in methods purely
based  on   maximum  likelihood,  the   results  are  biased   towards  faster
evolutionary   phases,    a   bias   known   as    the   'terminal-age   bias'
\citep{pont:2004}.

\begin{figure}
\begin{center}
\hbox{
\includegraphics[width=85mm]{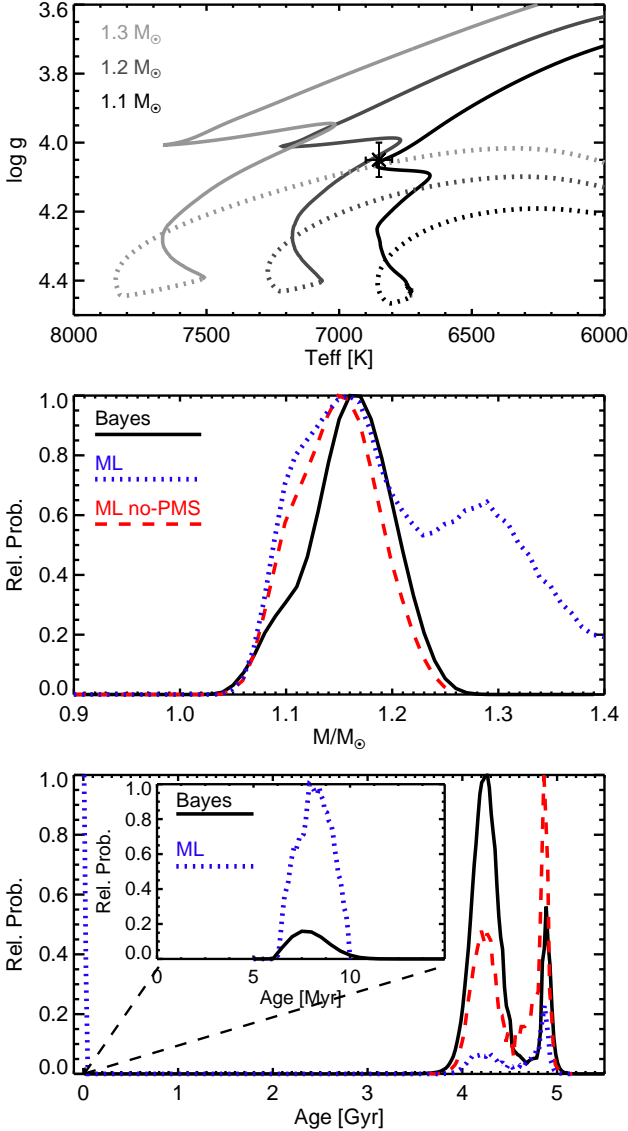}}
\caption{Top panel:  degeneracy of  evolutionary tracks on  the $\Teff -
    \logg$  plane at the  location of our  fiducial star (see  text). Middle
  panel:  relative probability  of  the mass  distribution functions  obtained
  using pure  maximum likelihood (ML  - blue dotted line),  maximum likelihood
  with suppressed pre-main sequence models  (ML no-PMS - red dashed line), and
  Bayesian analysis  (Bayes - black solid  line). Bottom panel:  same as above
  but  for the  age  distribution functions;  the  inset shows  in detail  the
  physically-disfavoured  narrow peak corresponding  to the  pre-main sequence
  solution (see text for details).
\label{fig:example_bayes}}
\end{center}
\end{figure}

To illustrate  here the importance of  using information not  contained in the
likelihood,  we present  a similar  example.   Let us  assume a  star has  the
following    observed     parameters:    \hbox{$\Teff=    6850\pm50$~K}    and
\hbox{$\logg=4.05\pm0.05$~dex}.   For simplicity,  we  exclude metallicity  in
this   example  by  fixing   it  to   [Fe/H] $=-1.0$.   In   the  top   panel 
of Fig.~\ref{fig:example_bayes} we show three  evolutionary tracks of models
with different masses overlapping  in the range of the  observed parameters. 
While the  1.1~M$_\odot$   model  is  entering   the  subgiant  branch   (SGB), 
the 1.2~M$_\odot$  is on the  main sequence  (MS) and  the 1.3~M$_\odot$  model
is still on the pre-main sequence (PMS).

First, we  perform a standard  maximum likelihood analysis (hereafter,  ML) to
determine the  stellar mass and age  based on Monte-Carlo  sampling. We create
synthetic  distributions  of $\Teff$  and  $\logg$ assuming  Gaussian
distributions of errors and search  for the best-fit model in all evolutionary
tracks of the  assumed metallicity.  The resulting mass  and age distributions
are shown  in the  middle and bottom  panels of  Fig. \ref{fig:example_bayes},
respectively.  The mass distribution shows  a bimodality with a secondary peak
and a tail  extending towards larger masses ($M>1.2~{\rm  M_\odot}$).  The age
distribution shows  a dominant  peak at very  low ages  (${\rm 6~Myr <  \tau <
  10~Myr}$) clearly seen in the inset  that zooms in in this age range.  There
are two other  local maxima at about 4.3~Gyr and 4.8~Gyr,  but with much lower
relative probabilities than the solution at  very low age. Thus the ML results
indicate a strong preference for the PMS solution for the observed parameters.

While  it  cannot be  completely  ruled  out that  the  star  is  on the  PMS,
considerations  of  the  shortness  of  this evolutionary  phase  render  this
possibility, a  priori, not very  likely. A 'brute  force' way of  taking this
into   consideration  is   to   directly  eliminate   PMS   models  from   the
analysis\footnote{This has  been done,  for example, in  asteroseismic studies
  \citet{gai:2011} because PMS models  have global oscillation properties that
  can  be  degenerate  with  those  of  MS  counterparts.}.   We  impose  this
restriction by removing all stellar models  with ages $\tau < 20$~Myr. In this
way, while the  best-fit model is still determined purely from  its ML, we are
already using prior information by assuming the chance of observing a PMS star
is   negligible   (actually   null,    although   this   also   is   a   wrong
assumption!). Results for this exercise are labelled 'ML no-PMS' in the middle
and bottom panels of  Fig.~\ref{fig:example_bayes}.  In the mass distribution,
the tail  at high  masses vanished since  its origin  was purely from  the PMS
models.  The  same is  true for  the ages: the  unrealistically large  peak at
$\sim 6$ Myr 
disappeared,  and  the peaks  at  4.3~Gyr and  4.8~Gyr  are  now the  dominant
features,  but their  relative weight  is still  the same  as in  the  pure ML
analysis  and the older  peak, associated  with stars  entering the  SGB still
dominates.

The results of  the Bayesian analysis for our fiducial star  are also shown in
Fig.  \ref{fig:example_bayes} (middle and bottom panels, solid black line). As
in  the  case  of the  ML  case,  three  peaks  are  present but  the  maximum
probability  in each  of them  reflects information  not contained  in  the ML
analysis.  The  PMS solution, although possible, is  now strongly disfavoured,
whereas the MS  solution is now the  most likely one and it  outweighs the SGB
solution. The Bayesian  mass distribution (middle panel) is  similar to the ML
no-PMS case. The  vanishing contribution of the PMS is  now easy to understand
from the eq.  \ref{eq:ageprob}.  The  probability for a given mass is obtained
by marginalizing the posterior PDF over all possible ages.  The stellar tracks
with high masses, which contribute in  the observational box with the PMS, are
strongly disfavoured  because of the  short duration of this  phase.  However,
the  PMS  solution   still  has  a  non-negligible  probability   in  the  age
distribution.   The widths  of the  peaks reflect  evolutionary speed  in each
phase,  with  PMS  evolution being  about  $100$  times  faster than  late  MS
evolution and early SGB a few times faster than late MS.

There is  one important aspect related  to the interpetation of  the height of
the peaks  in the Bayesian age  PDF.  Strictly speaking,  when determining the
probability of a given age $p(\tau)$ using Equation~\ref{eq:ageprob}, there is
no information on  the speed of stellar evolution because we  are looking at a
time snapshot.   As stated by  \cite{jorgensen:2005}, the relative  heights of
the  peaks  in the  age  distribution are  determined  by  how closely  packed
evolutionary tracks of different masses are around the observational error box
(i.e., the  region over  which likelihood is  significant and  varies little).
Where variations in stellar mass introduce  small changes on the location of a
stellar model  (for the same age), the  integrand in Equation~\ref{eq:ageprob}
will vary  slowly when integrating in  mass.  Thus, the  interpretation of the
height of  the peaks as  depending on the  evolutionary speed of stars  is not
straightforward.  It  only follows if slow evolutionary  speeds are associated
with regions  on the observational plane  where small changes  in mass produce
small changes  in the observational  quantities the determine  the likelihood.
As stated above, the unique direct manifestation of a slowly evolving phase in
the age PDF is the width of the peaks in the distribution.

\subsection{Examples} \label{sec:realstars}

\begin{figure}
\begin{center}
\includegraphics[width=85mm]{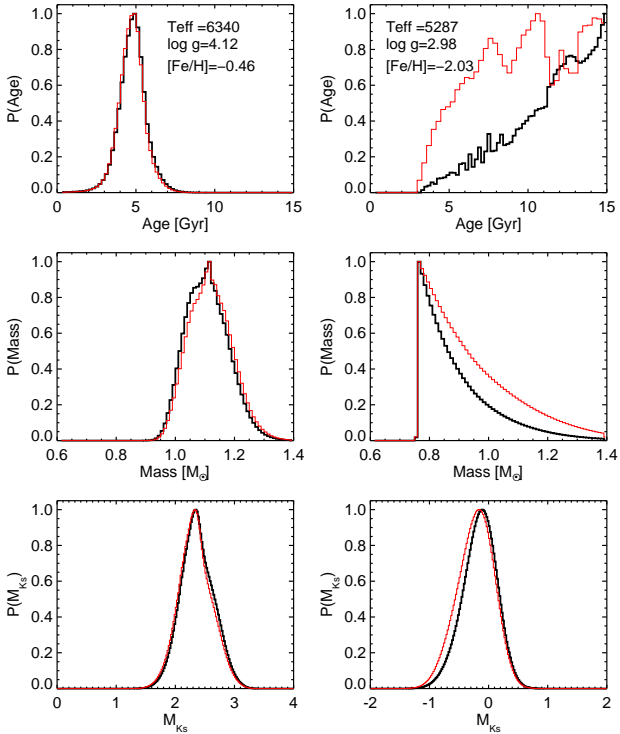}
\caption{Probability distribution functions (PDFs) for age, mass, and absolute
  magnitude in  the ${\rm  K_S}$ band.  On  the left  results are shown  for a
  turn-off star and, on  the right for a star on the  red giant branch. In the
  latter case, neither  mass or age can be  determined. The absolute magnitude
  (luminosity),  on  the  other  hand,   can  be  derived  precisely  in  both
  cases. Black solid lines: Salpeter IMF  as a prior in the mass distribution;
  red dashed lines: flat IFM.
\label{fig:pdf}}
\end{center}
\end{figure}

Fig.~\ref{fig:pdf} shows in  black thick lines the PDFs  of stellar parameters
of two  stars from  our sample.  On the left,  J112344.9-290327 has  [Fe/H] $=
-0.46 \pm 0.10$, $\Teff=6340\pm70$~K, and $\logg= 4.12\pm0.10$, i.e. it
is  a  moderately metal-poor  turn-off  star.  In  this  case,  mass, age  and
absolute $Ks$ magnitude\footnote{Hereafter, we  refer to 2MASS $Ks$ magnitudes
  only, because they are used to determine distances.}  have PDFs that lead to
well    determined    central     values    and    uncertainties.    Following
\cite{burnett:2010},  we  compute  stellar  parameters and  their  uncertainty
directly from  the different  moments of  the PDFs.  In  this case,  we obtain
$\tau=4.78 \pm0.88$~Gyr, ${\rm  M=1.108\pm 0.071~M_\odot}$, and ${\rm M_{Ks}}=
2.371 \pm 0.279$.

As a second  case we have selected J115337.3-020036, with  [Fe/H] $= -2.03 \pm
0.10$, $\Teff=5287\pm68$~K, and  $\logg= 2.98\pm0.10$,  a star  on the
lower part of the RGB. As it  is well known, evolutionary tracks of stars with
different masses and  similar metallicity are almost degenerate  along the RGB
in  the  $\Teff  -  \logg$  plane (Fig.~\ref{fig:tracks_data}).   This
degeneracy precludes  the determination of its  mass and age as  seen from the
PDFs in Fig.~\ref{fig:pdf}.  The cut  in the age distribution at 15~Gyr simply
reflects our  age prior, and this corresponds  to the mass cut  at $\sim 0.75$
M$_\odot$.  The shapes of  the PDFs  are almost  completely determined  by the
adopted priors,  i.e., IMF and  SFR, and by  the fact that more  massive stars
evolve faster along the RGB than lower mass stars. No actual information about
the true stellar  mass and age can be  obtained in this case. In  fact, a pure
likelihood analysis  leads to  age and mass  distributions that  are basically
flat.   On the  other  hand, the  absolute  magnitude of  this  star is  still
well-defined, and  in fact yields ${\rm  M_{Ks}}= -0.155 \pm  0.289$, the same
precision as  for the  turn-off star.  This  conclusion remains valid  for all
stars along the RGB, the absolute magnitude is a well-defined quantity even if
the underlying evolutionary tracks are  highly degenerate, and we can robustly
estimate  stellar distances  independently of  the evolutionary  stage  of the
stars.

Finally, we have tested the impact of the IMF prior by considering also a flat
IMF.   Results  for   the   two   examples  just   discussed   are  shown   in
Figure~\ref{fig:pdf} in red dashed line. Clearly, the impact of changing the IMF
is minimal  in those cases where  stellar parameters are well  defined. On the
contrary, the large impact of priors in the age and mass distributions for the
giant star are  evidence that no actual information can  be extracted in these
cases: the posterior PDF is largely determined by the priors.

%
%

\section{Results}{\label{sec:results}}

Using  the procedure  described in  Sect.~\ref{sec:method} and  the  two sets,
$\pone$ and $\ptwo$ of stellar parameters (Section~\ref{sec:obs}), we computed
ages,  masses  and  absolute magnitudes  for  the  stars  in our  sample.   To
determine the distances, we used the observed 2MASS ${\rm K_S}$ magnitudes and
the absolute ${\rm K_S}$ magnitudes  obtained from the Bayesian analysis.  The
synthetic magnitudes  were computed from  the MARCS grid of  synthetic spectra
\citep{gustafsson:2008} assuming plane-parallel  radiative transfer for $\logg
> 3$  and spherically-symmetric  mode  otherwise.  The  synthetic ${\rm  K_S}$
magnitudes  have   been  computed  adopting  the   absolute  calibrations  and
zero-points discussed in \citet{casagrande:2010}.  Reddening for each star was
computed  using \citet{schlegel:1998} maps  following the  iterative procedure
described in \citet{ruchti:2011}.

\begin{figure}
\begin{center}
\includegraphics[width=85mm]{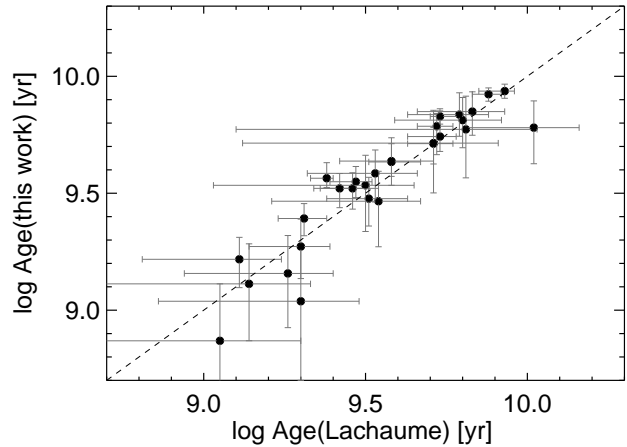}
\caption{Comparison of ages determined for the sample of nearby stars of 
\citet{lachaume:1999}. 
\label{fig:lach}}
\end{center}
\end{figure}

\begin{figure}
\includegraphics[width=85mm]{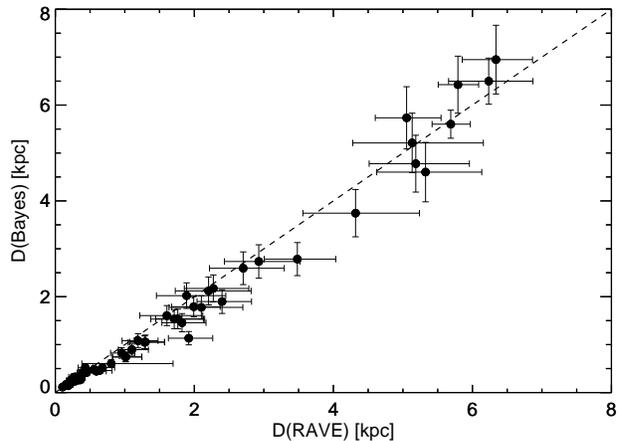}
\caption{Comparison  of  distance  determination  methods. We  show  distances
  obtained  with our  Bayesian  scheme  and those  by  the RAVE  collaboration
  \citep{zwitter:2010}.   Note that,  for this  test only,  the  input stellar
  parameters ($\Teff$,  $\logg$, [Fe/H]) are  from the RAVE DR3  catalogue for
  both distance determinations. 
\label{fig:valid_dist}}
\end{figure}

\subsection{Tests}

In order  to assess  the robustness  of our Bayesian  scheme, we  selected two
sub-samples of stars with  basic stellar parameters from \citet{lachaume:1999}
and from \citet{zwitter:2010}.  For the former, we determine  ages and compare
them with the results from \citet{lachaume:1999}.  For the latter, we focus on
the distances.

For   stellar  ages,   the   comparison  only   comprises   stars  for   which
\citet{lachaume:1999}   derived   ages   from  isochrone   fitting\footnote{In
  addition, stars hotter than $\sim  7500~$K were excluded from the comparison
  because the  grid of  stellar evolution models  is not complete  above those
  temperatures  and  because  we  use  the  \texttt{MARCS}  model  atmospheres
  \citep{gustafsson:2008}, which are available for  $\Teff < 8000$ K only.} in
order to  avoid biases derived from  comparing ages obtained  with methods not
based   on   stellar   evolutionary   tracks.    The   very   good   agreement
(Fig.~\ref{fig:lach})  implies that when  the observational  uncertainties are
small, $1\%$ in $\Teff$ and  $0.05$ mag in $M_{v}$, maximum likelihood methods
return similar  results to  the Bayesian analysis  and age  determinations are
robust.   A similar  agreement  was reported  by \citet{jorgensen:2005}.   The
uncertainties in  the ages  we derive are  generally smaller compared  both to
\citet{lachaume:1999} and \citet{jorgensen:2005}.

We have derived stellar distances using the input stellar parameters from 
\citet{zwitter:2010}\footnote {We 
  used the  RAVE distances obtained  for stars with repeated  observations and
  Padova  isochrones.  This combination  should  provide reliable  estimations
  according to  the note added in  proof in \citet{zwitter:2010}.}  for the 58
stars common with our sample, and show the results in 
comparison      with      RAVE      distances     \citep{zwitter:2010}      in
Fig.~\ref{fig:valid_dist}. Results are an excellent agreement.
Therefore, we conclude that, provided the same input stellar
parameters are used,  our Bayesian scheme and the isochrone matching
method by 
\citet{zwitter:2010} yield  very similar  results, with no  obvious systematic
offsets between methods.  This is in line with  \citet{burnett:2011}, who also
applied a  (more sophisticated) Bayesian  analysis to the RAVE  stellar sample
and found that,  although some systematic differences were  found, the overall
agreement with \citet{zwitter:2010} was good. 

\subsection{Ages and Masses}
\begin{figure}
\begin{center}
\includegraphics[width=85mm]{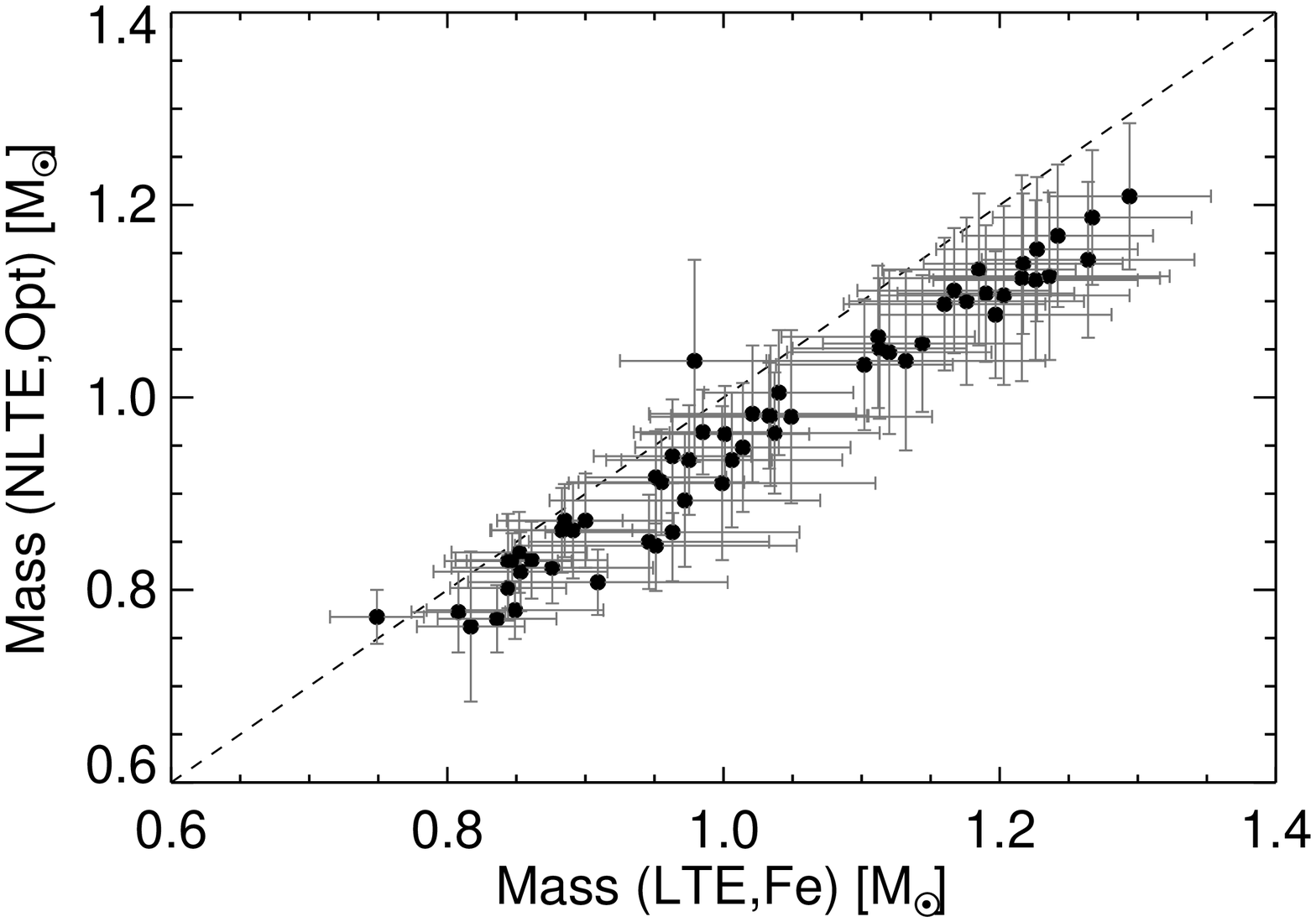}
\includegraphics[width=85mm]{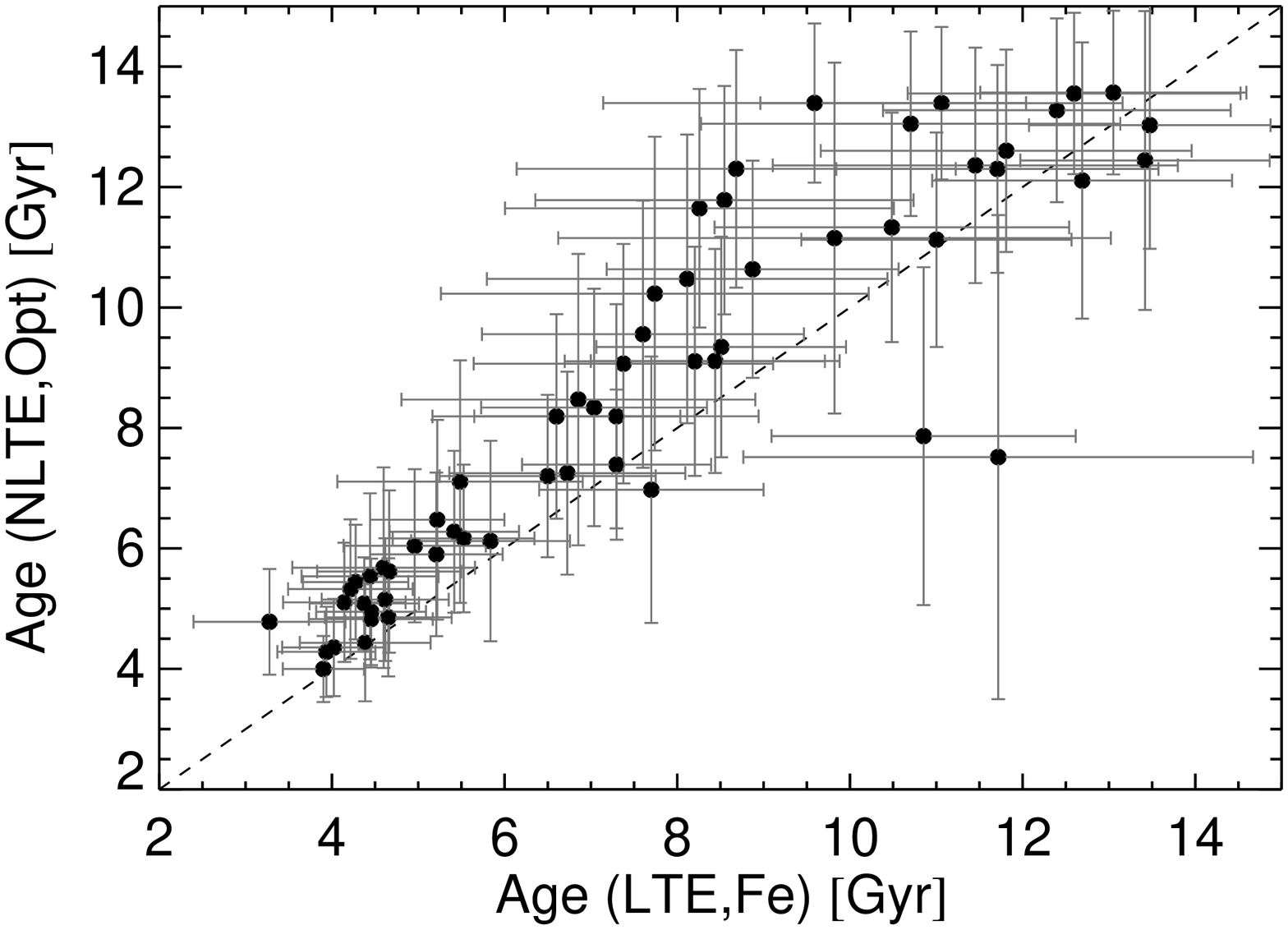}
\caption{Masses and ages for our sample of stars derived with the NLTE-Opt and
LTE-Fe basic stellar parameters. Only stars with $\logg > 3.6$ are
included (see text). 
\label{fig:agemass}}
\end{center}
\end{figure}
Having verified  the reliability of the  Bayesian method, we  proceed with the
analysis of our stellar sample.  The degeneracy of stellar tracks discussed in
Sec.~\ref{sec:realstars} limits the ability to  derive masses and
  ages on the RGB. For this reason, we derived masses and ages for stars with
$\logg \ge 3.6$, that is, for $59$ stars only.

In Fig.~\ref{fig:agemass} we compare masses and ages derived from the two sets
$\pone$ and $\ptwo$  of input stellar parameters. Stellar  masses derived from
the  $\ptwo$  parameters are  systematically  lower  than  those derived  from
$\pone$,  by  about  $0.05  -  0.1$  M$_\odot$. The  primary  reason  for  the
difference   lies  in  the   larger  gravities   obtained  with   the  $\ptwo$
method. Errors of  the masses are comparable in both  cases, and range between
0.03~M$_\odot$  up to  0.10~M$_\odot$. Nevertheless,  the systematic  shift to
lower masses caused by the use of more accurate $\ptwo$ stellar parameters can
not be neglected.  

For stellar ages, the behavior is opposite, as expected.  The results based on
the $\ptwo$ parameters are higher by $10$ to $30$ percent. However, not in all
cases a smaller mass implies a  larger age. Close inspection of both panels in
Fig.~\ref{fig:agemass}  reveals only  two  stars with  larger $\ptwo$  masses,
whereas six of them have smaller ages derived from $\ptwo$ parameters. Typical
uncertainties of  our Bayesian ages are of  the order of 20\%  and, except for
two cases,  range between 10\%  and 30\%. We  conclude that the  LTE-Fe method
underestimates stellar ages by $1 - 4$~Gyr, and the errors are more pronounced
for hotter and more metal-poor stars.

\subsection{Distances}

The main goal  of this section is  to show the dramatic impact  that using the
physically realistic  $\ptwo$ stellar parameters  has on the  determination of
stellar distances.

\begin{figure}
\begin{center}
\includegraphics[width=85mm]{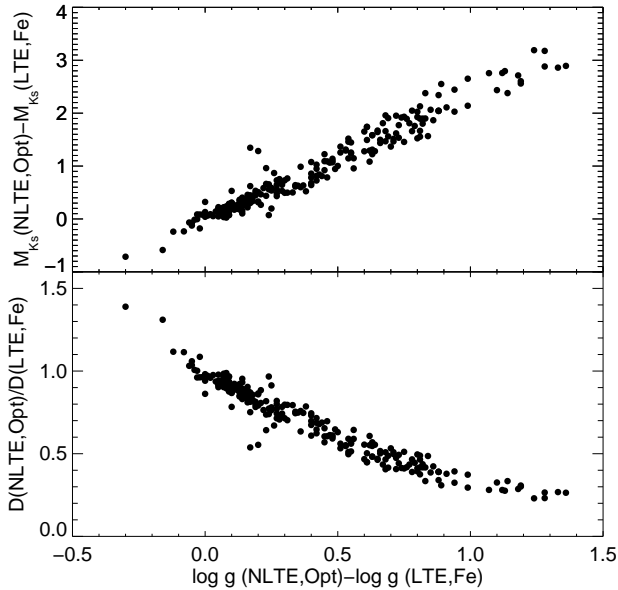}
\caption{Changes  in the  absolute  $K_S$  magnitudes  and   distances
  caused  by  using  the  improved  set of  input  basic  stellar  parameters,
  $\ptwo$ as a function of the change in $\logg$, for our full sample.
\label{dist_rel}}
\end{center}
\end{figure}

As summarized in Sect.~\ref{sec:obs},  there are large differences between the
$\pone$   and   $\ptwo$   stellar   parameters   sets.  The   top   panel   of
Fig.~\ref{dist_rel} shows the change in the absolute magnitude ${\rm M_{K_S}}$
as  a  function  of  the  $\gtwo  -  \gone$  differences.  The  strong  linear
correlation simply reflects the fact  that changes in $\logg$ are accompanied
by  a  reciprocal  change  in  luminosity  due  to  their  inverse  functional
dependence  on  stellar  radius.   Clearly,  the  $\ptwo$  stellar  parameters
indicate that  all our stars,  with few exceptions, are  \textit{fainter} than
what would be inferred from using the LTE-Fe parameters.  Furthermore, if, for
the  sake of discussion,  we neglect  the effect  of reddening,  then distance
scales exactly as the square root of the luminosity of the star. Consequently,
distances derived  from the  $\pone$ set of  parameters are  overestimated, as
illustrated in Fig.~\ref{dist_rel}  (bottom panel) where we show  the ratio of
distances derived using $\pone$ and $\ptwo$ stellar parameters.

\begin{figure}
\includegraphics[width=85mm]{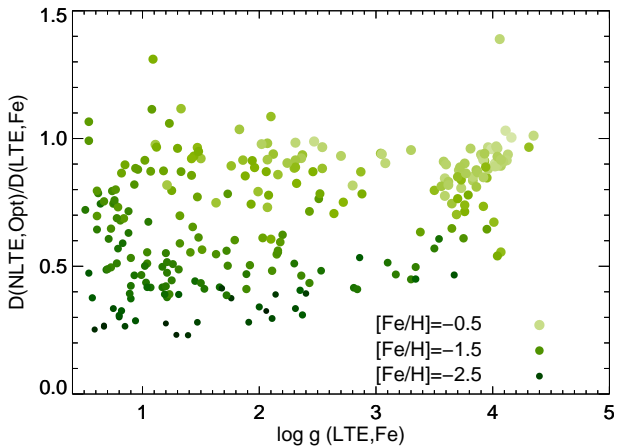}
\caption{Ratio of distances derived with $\pone$ and $\ptwo$ stellar parameters
as a function of surface gravity for our full sample. Symbol sizes and colours
indicate the stellar metallicity. 
\label{fig:dist_logg}}
\end{figure}

Similar information is provided in Fig.~\ref{fig:dist_logg} in which the ratio
of distances, corrected for reddening, is plotted as a function of 
$\log g_{\rm LTE,Fe}$.  This illustrates clearly how systematic errors in
distance
  determinations  impact different  stellar types.  Stellar  metallicities are
  encoded   in  the   symbol  sizes   and   colours,  as   indicated  in   the
  legend. Distances to  metal-rich stars, [Fe/H] $> -1$,  are, to first order,
  unbiased.  On the  other hand, for metal-poor ([Fe/H]  $\leq -1$) dwarfs and
  sub-giants ($\logg  > 3.6$), LTE-Fe  parameters imply a  systematic distance
  error of about $10 - 40\%$.  The problems are exacerbated at low metallicity
  and  low gravity.   Metal-poor  giants suffer  from  the largest  fractional
  distance biases of up to $70\%$.

\begin{figure}
\includegraphics[width=85mm]{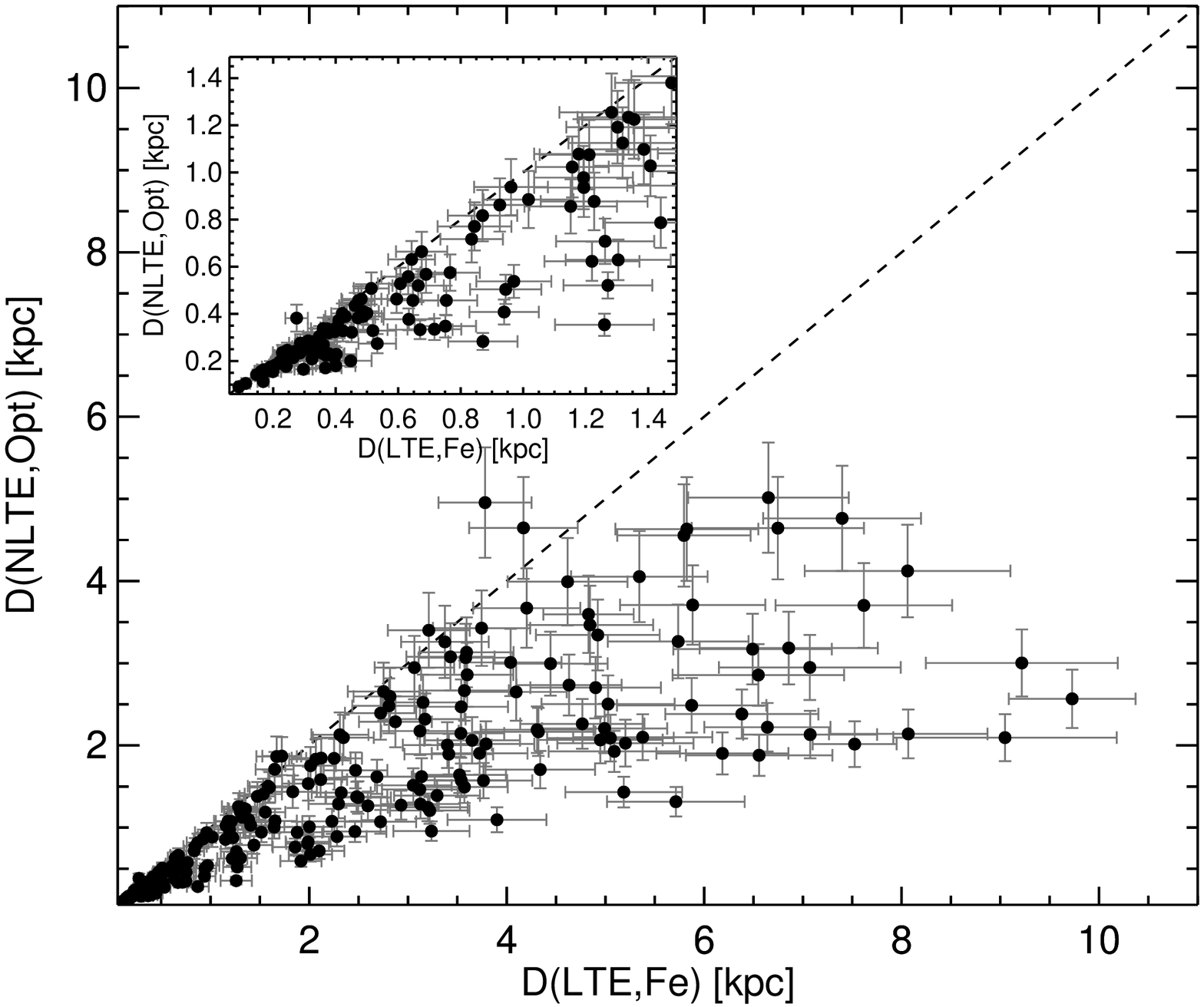}
\caption{Comparison  of heliocentric distances  derived for  all stars  in our
  sample  using $\pone$  and  $\ptwo$  sets of  basic  stellar parameters.  In
  virtually  all cases,  $\pone$ parameters  lead to  large  overestimation of
  distances.
\label{fig:dist_abs}}
\includegraphics[width=85mm]{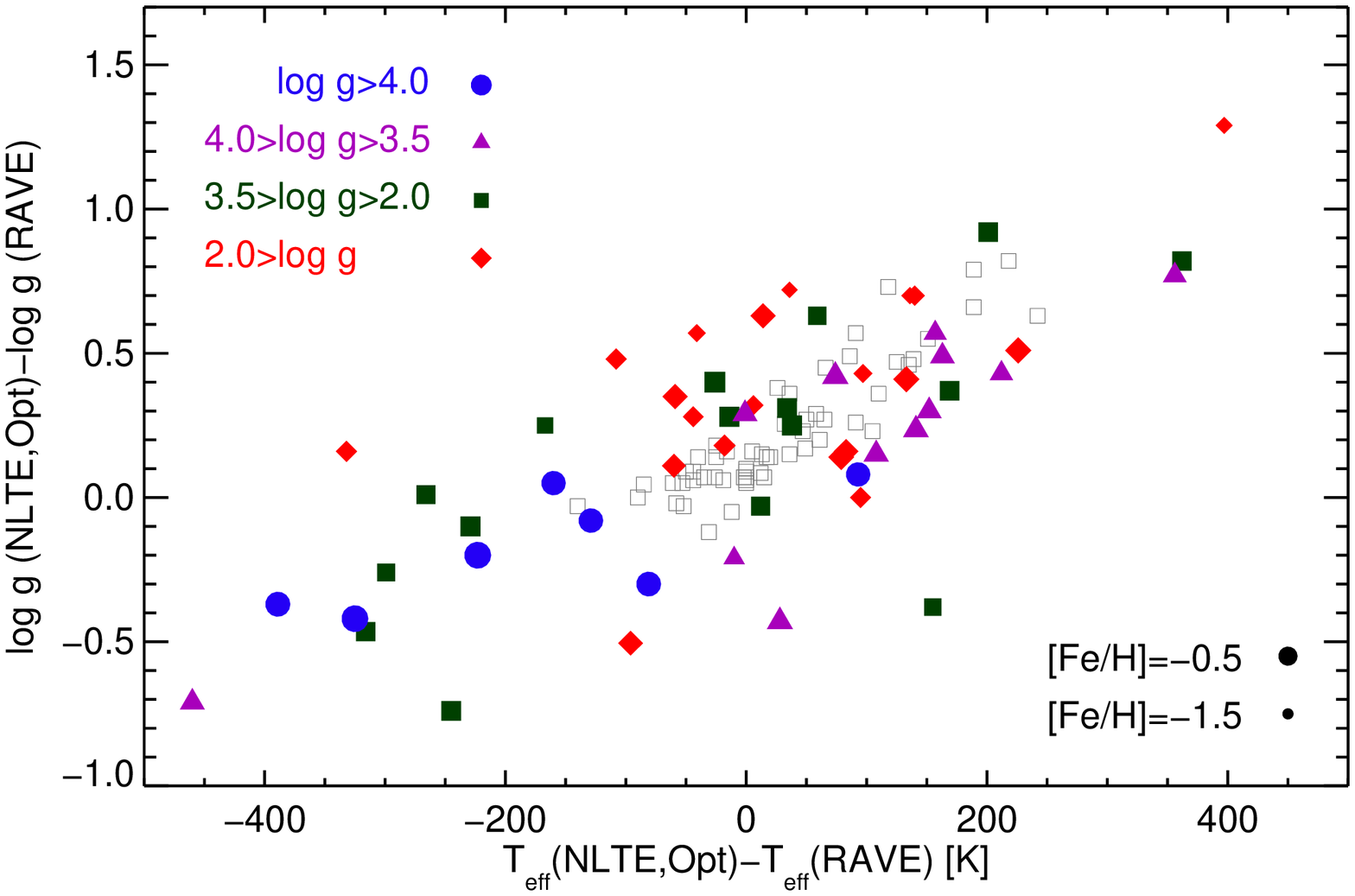}
\caption{Differences  in basic  stellar  parameters for  the  stars in  common
  between     our    sample     and    the     RAVE     distances    catalogue
  \citep{zwitter:2010}.  Colours and  symbol types  are assigned  according to
  $\logg$ values from the  DR3 pipeline. Symbol sizes are indicative of
  the DR3  metallicity, as indicated  in the legend.  The  differences between
  (NLTE,Opt)  and (LTE,Fe) parameters  for the  same stars  are shown  in grey
  empty squares.
\label{fig:rave_vs_us_para}} 
\includegraphics[width=85mm,height=53mm]{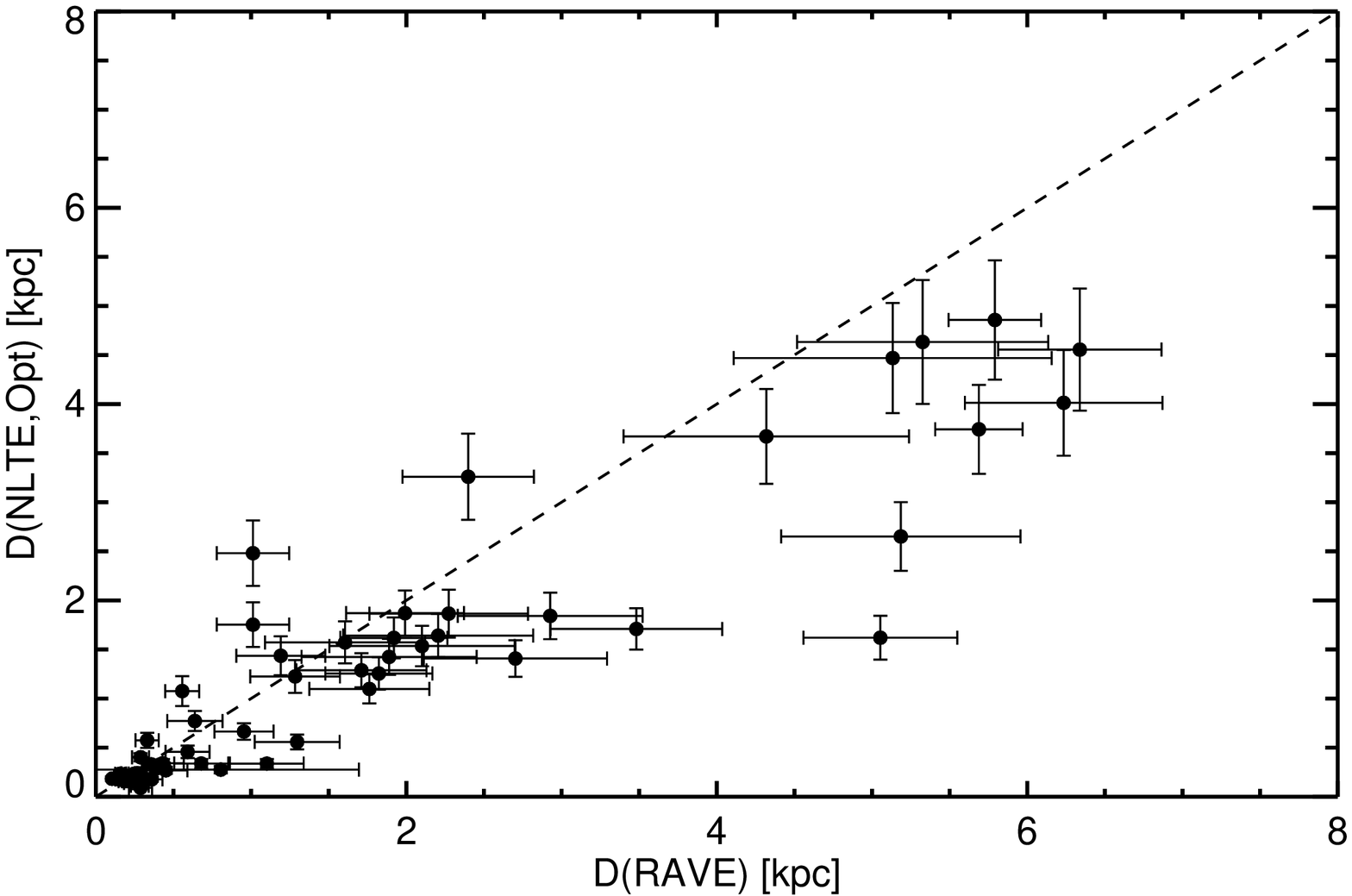}
\caption{Comparison of distances derived using $\ptwo$ stellar parameters and 
those from the RAVE DR3 catalogue for stars common in both samples. The 
trend strongly resembles that of the comparison between $\pone$ and $\ptwo$ 
distances (Fig.~\ref{fig:dist_abs}).
\label{fig:rave_vs_us_dist}}
\end{figure}

Fig.~\ref{fig:dist_abs} shows  the final distribution  of distances determined
using the  $\ptwo$ and  $\pone$ stellar parameters  for our full  sample. This
plot summarizes  what has already been  discussed above, and allows  us to put
the results into the more  general perspective of large-scale stellar surveys.
In a  magnitude-limited survey, for  example, where more  metal-rich unevolved
stars  dominate   the  nearby  sample  and  metal-poor   luminous  giants  are
predominantly   observed   at   large   distances,   classical   full   LTE-Fe
analysis\footnote{It is important  to remind the reader that  LTE-Fe refers to
  the classical  spectroscopic method of LTE  excitation-ionization balance of
  Fe, which is very efficient because  $\Teff$, $\logg$, and [Fe/H] can be all
  determined  simultaneously.}   will  systematically overestimate  distances,
placing  stars  progressively  further  than  they are.   This  has  important
consequences  in   all  cases  when  stellar   metallicities,  positions,  and
kinematics  are   used  for  studies  of  Galactic   structure  and  evolution
\citep{siebert:2008, carollo:2010,  wilson:2011, bensby:2011}.  We  touch upon
this in Sect.~\ref{sec:summary}.

\subsection{Comparison      with     other      studies:      RAVE     stellar
  parameters}\label{sec:ravecomp} 

Since our  sample of stars was drawn  from the internal RAVE  catalogue, it is
very important to investigate how  our results, i.e.  $\Teff$, $\logg$,
[Fe/H], and distances, compare with  those obtained by the RAVE collaboration.
In   what    follows,   we   will    compare   our   data   with    those   by
\citet{zwitter:2010}\footnote{\citet{seibert:2011} have  presented the revised
  RAVE DR3 pipeline. However $\Teff$ and $\logg$ for the 58 stars
  we use in  this comparison are exactly the  same as in \citet{zwitter:2010};
  only [Fe/H] differ by about 0.1 to 0.2~dex, and this has a very minor impact
  on the  determination of  distances.  We  thus prefer to  use all  data from
  \citet{zwitter:2010},  which has been  used by  RAVE to  obtain distances.},
with which we  have $58$ stars in common. We  highlight the key methodological
differences below.
\begin{itemize} 
 \item  The quality of  observations. Our  parameters rely  on high-resolution
   data, R $\sim 30000$, with a  very broad wavelength coverage, $3500 - 9500$
   \AA;  the  RAVE  spectra have  R  $\sim  7500$  and  cover only  a  limited
   wavelength range between $8410$ and $8795$ \AA.
 \item The  physics of stellar  atmospheres and radiative transfer  models. In
   particular, our analysis relies on  NLTE radiative transfer, while the RAVE
   spectroscopic pipeline operates fully on 1D LTE synthetic spectra.
 \item  The different  approaches to  determine masses  and ages,  and stellar
   evolution models.  \texttt{GARSTEC} tracks  were used in this work, whereas
   \citet{zwitter:2010} adopted Padova isochrones. Also, the approach followed
   in that study is esentially  a nearest neighbor technique, however with the
   important  difference  that the  isochrones  were  resampled  to a  uniform
   spacing in age to account for the variable speed of stellar evolution, and 
   a weighing scheme to include, for example, the impact of the initial mass 
   function. In this  respect, their  analysis also makes use of priors.
 \item   \citet{zwitter:2010}   neglected    reddening   in   their   distance
   determinations.    We    used   the   reddening    values following
   \citet{ruchti:2011}; also comparison with the interstellar NaD lines in our
   high-resolution spectra indicates that reddening is not negligible for many
   stars.
 \item The uncertainties on our input stellar parameters are typically smaller
   than  those of  \citet{zwitter:2010}. Their  values are  $\sim 0.2$  dex in
   [Fe/H] and $\logg$ and range between $\sim 100$ to 200~K in most cases for
   $\Teff$,  whereas our standard  estimates are $60-100$ K  in $\Teff$,
   and $0.1$ dex in [Fe/H] and $\logg$.
\end{itemize}

The  differences  between $\ptwo$  parameters  and  those  from the  RAVE  DR3
pipeline are  shown in Fig.~\ref{fig:rave_vs_us_para}. Size and  colour of the
symbols  indicate  metallicity  and  gravity  (we used  RAVE  DR3  values  for
reference).  We also show  the differences  $\ptwo$ and  $\pone$ for  the same
stars in grey empty squares. RAVE DR3 parameters show a very large spread with
respect to  $\ptwo$. In fact, the spread  is much larger than  with respect to
the  $\pone$ parameters. Clearly  it is  not possible  to single  out specific
factors which  cause the  spread; however, some  conclusions can be  drawn. In
most cases, the RAVE DR3  $\logg$ values are clearly underestimated for giants
with  $\Delta \logg_{(\rm{NLTE,Opt -RAVE~DR3})}$  up to  $\sim 1$~dex  and one
case as  large as 1.3~dex. This  effect is more pronounced  for the metal-poor
stars,   in   qualitative  agreement   with   the   differences  between   the
$\ptwo$-$\pone$ parameters for the same stars. The same is true for subgiants;
in most cases  the DR3 gravities are underestimated. For  dwarfs, on the other
hand,   we   find   DR3   gravities   are  overestimated   by   as   much   as
0.5~dex.  \citet{burnett:2011} arrive  at  the same  conclusion comparing  DR3
gravities with stellar models.

As  a  consequence of  the  differences  in  stellar parameters,  particularly
changes in $\logg$, the distances we obtain from the $\ptwo$ parameters
are    systematically    \textit{smaller}    than    the   RAVE    DR3    data
(\citealt{zwitter:2010,  seibert:2011}).  For   the  most  distant  metal-poor
giants,  the offset  is of  the order  $50$ percent.   This is  illustrated in
Figure~\ref{fig:rave_vs_us_dist}, which shows  that beyond 1.5~kpc all $\ptwo$
distances but one are smaller  than those derived in \citet{zwitter:2010}. For
a few dwarfs in the immediate solar neighborhood, the results are opposite, as
expected from the sign of the $\logg$ corrections. The overall trend is
very clear and the qualitative  similarity with the difference between $\ptwo$
and $\pone$ distances (Fig.~\ref{fig:dist_abs}) is striking.  

\begin{figure}
\centerline{\includegraphics[width=55mm, angle=-90]{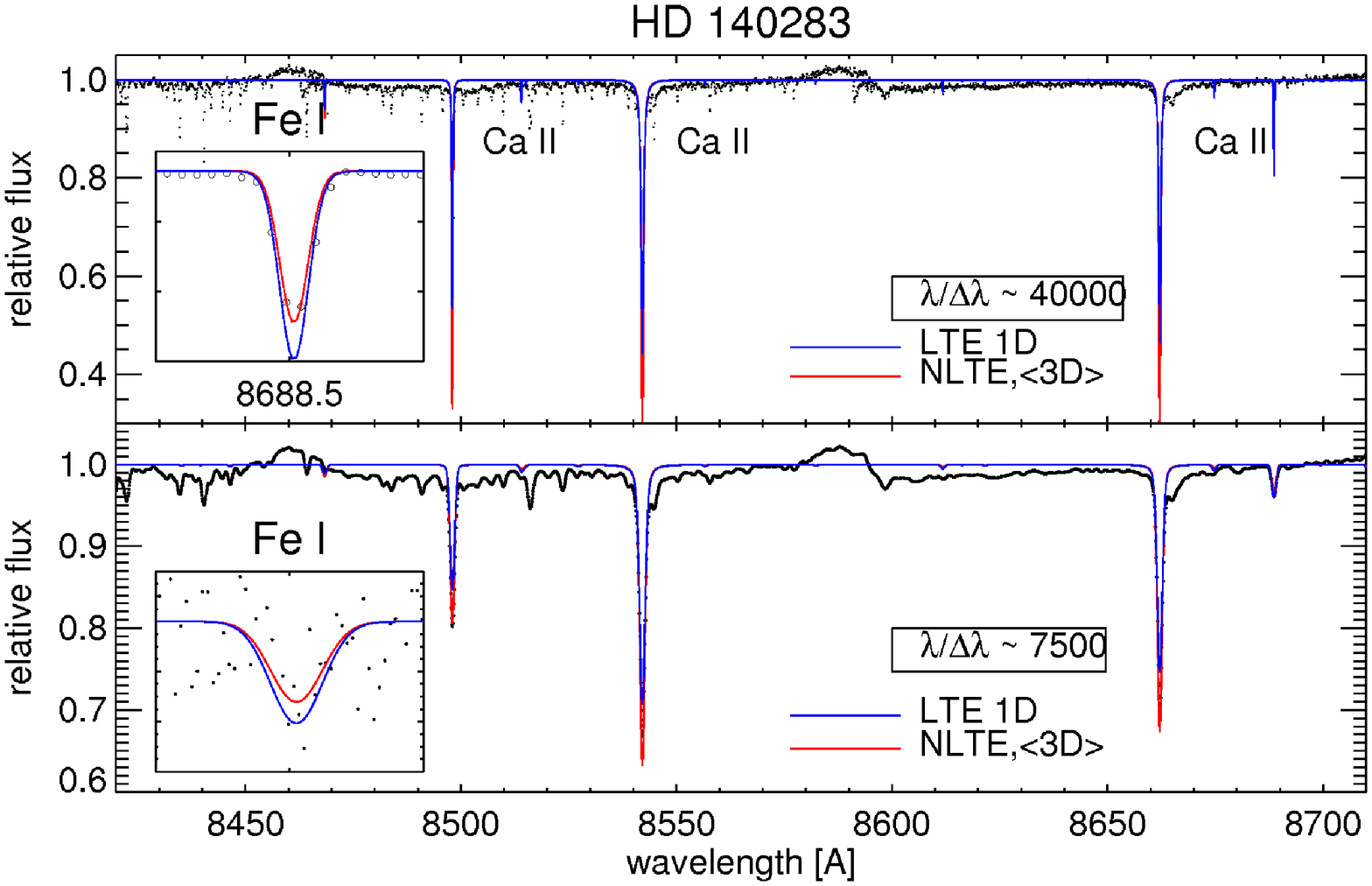}}
\centerline{\includegraphics[width=55mm, angle=-90]{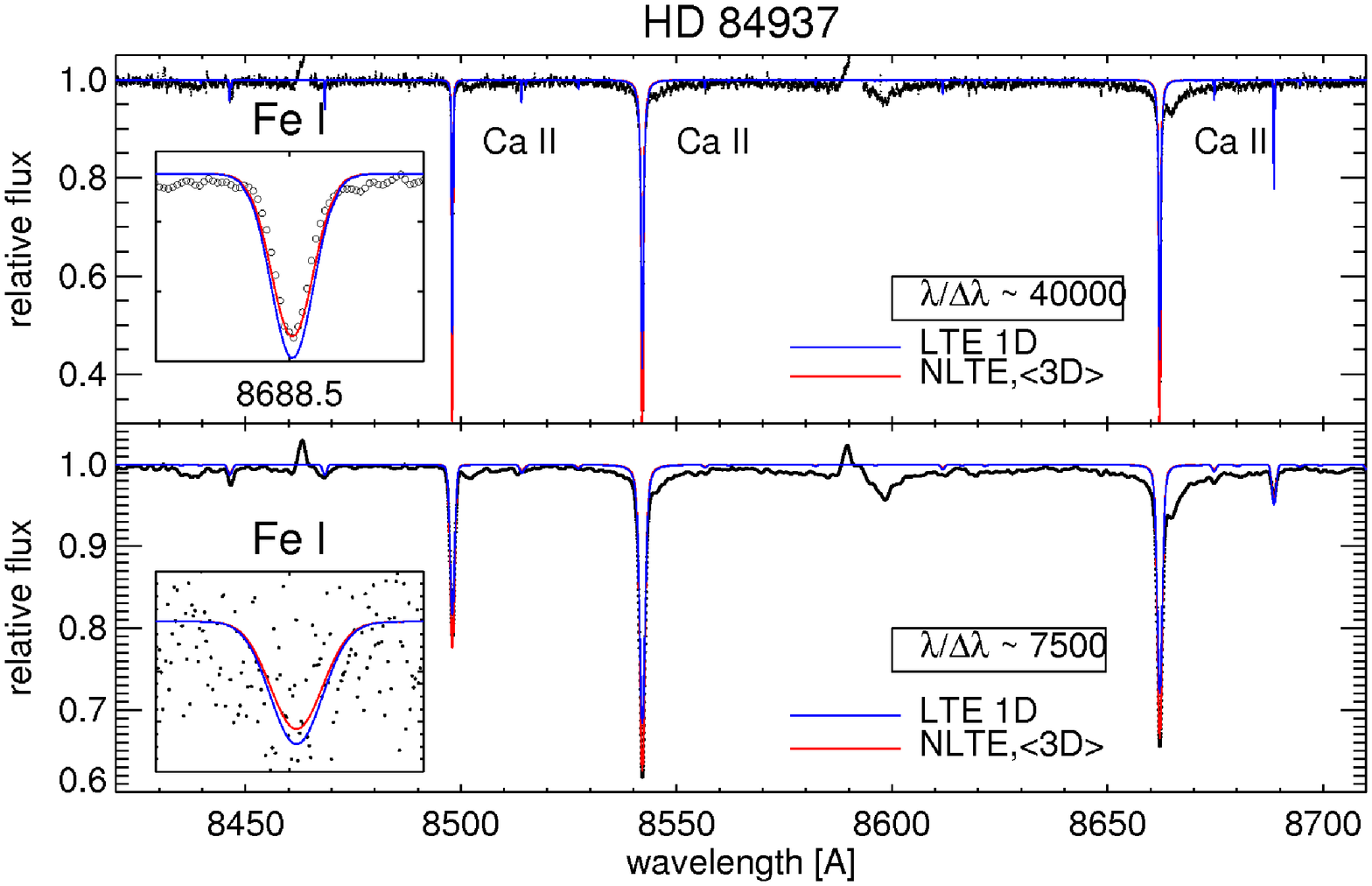}}
\caption{Synthetic 1D  LTE (blue) and  $<$3D$>$ NLTE (red)  spectra of
    the metal-poor stars HD 140283 (top  plot) and HD 84937 (bottom plot). The
    observed spectra  (black points)  are from Bergemann  \& Gehren  (2008) in
    their original resolution, $R \sim  40\,000$ (top panel in each plot), and
    degraded to  the resolution  of the RAVE  spectra $R \sim  7\,500$ (bottom
    panel in each  plot). The 1D LTE radiative  transfer models under-estimate
    the strength of the  Ca II lines and over-estimate that of  the Fe I lines
    (shown in the inset). See text for more details.
\label{fig:spectra}}
\end{figure}

These results provide  alarming evidence that the RAVE  DR3 stellar parameters
are  affected by  the systematic  error  caused by  the 1D  LTE assumption  in
spectroscopic  parameter  determinations.  Test  calculations  for the  Ca{II}
near-IR    triplet    using   the    mean    3D    model   atmospheres    from
\citet{bergemann:2012}  indicate  that   these  spectral  lines  are  strongly
affected by the ${\rm 1D-\left<3D\right>}$ differences, while the near-IR Fe I
lines are  mostly affected  by NLTE. Fig.~\ref{fig:spectra}  illustrates these
effects for the  two well-studied metal-poor stars, a  subgiant HD 140283 (top
plot) and a warm turn-off star  HD 84937 (bottom plot) with [Fe/H] $=-2.4$ and
$-2$ dex, respectively.  The spectra are from \citet{bergemann:2008} and shown
in the  top panels  in their  original resolution, $R  \sim 40\,000$.   In the
bottom  panels the  spectra have  been  degraded to  the quality  of the  RAVE
spectra,   i.e.     $R   \sim   7\,500$    and   a   typical    S/N   $\sim50$
\citep[][Fig.~8]{zwitter:2008}.   The $\left<{\rm  3D}\right>$  NLTE synthetic
spectra were computed as in \citet{bergemann:2012}, but applying LTE radiative
transfer to Ca{II}\footnote{\citet{mashonkina:2007} found that NLTE effects in
  the  near-IR Ca lines  (3d-3p multiplet)  are also  important, and  they may
  further strengthen  the lines  at low metallicity.}.   The 1D  LTE radiative
transfer models under-estimate the strength of the Ca{II} lines (the effect of
$\left<{\rm 3D}\right>$  structure) and over-estimate  that of the Fe  I lines
(the effect of NLTE).  Also, the  near-IR lines of $\alpha$-elements (Si I, Mg
I,  and  Ti  I), which  along  with  Fe  I  lines  dominate the  RAVE  spectra
\citep{zwitter:2008}   all   form    in   NLTE   \citep{zhao:2000,   shi:2011,
  bergemann:2012b}  and it is  unlikely that  the NLTE  effects cancel  out to
produce  a   spectrum  close  to   LTE.   The  effects  are   clearly  present
irrespectively of the spectral resolution.

We  close this  section  with a  few  comments related  to  the two  different
validations  that \citet{zwitter:2010}  presented  for distance  determination
from RAVE data.  The first one is a comparison with Hipparcos stars present in
their catalogue with trigonometric  errors smaller than 20\% and spectroscopic
distance  errors smaller than  32\%. These  results are  shown in  Figure~3 in
\citet{zwitter:2010} and a large spread  is present in the comparison.  Even
if no  systematic difference was  found, as noted  by the authors,  almost all
these  stars dwarfs with  a solar-like  metallicity ([M/H]=$-0.12  \pm 0.25$),
exactly the case  where we do not expect large NLTE  effects.  The second test
consisted in comparing about 14 stars in 7 open clusters with known distances.
Again, all these clusters have solar-like metallicity \citep{dias:2002} and we
do not expect NLTE  effects to play a central role. In  any case, we note that
distances derived by \citet[][see their Table~2]{zwitter:2010} for the cluster
stars deviate  from the  cluster distances by  anything between -20  to +40\%.
These stars  were reanalyzed by \citet[][see  their Table~3]{burnett:2011} and
results show a similar spread, although in a few cases distance determinations
are also quite different from those in \citet{zwitter:2010}. It is simply not 
possible  to   try  to  establish  confidence   in  spectroscopically  derived
parameters, particularly at low metallicities and gravities, by extrapolating
confidence  gained by  means of  tests  performed in  very different  regimes,
e.g. at solar metallicity. 
%
%
%
\section{Summary and conclusions}\label{sec:summary}

This paper is the second in the series where we explore the influence of
different  methods  on the  determination  of  basic  and fundamental  stellar
parameters for FGKM  stars. The main goal is to understand  to what extent our
knowledge  about stars,  their surface  and interior  structure, is  biased by
inferences  derived from  classical model  atmospheres with  their simplifying
assumptions (1D  hydrostatic and local thermodynamic  equilibrium, LTE), which
have been used as a basis for spectroscopic characterization of stars over the
past $70$ years.

\begin{figure}
\includegraphics[width=90mm]{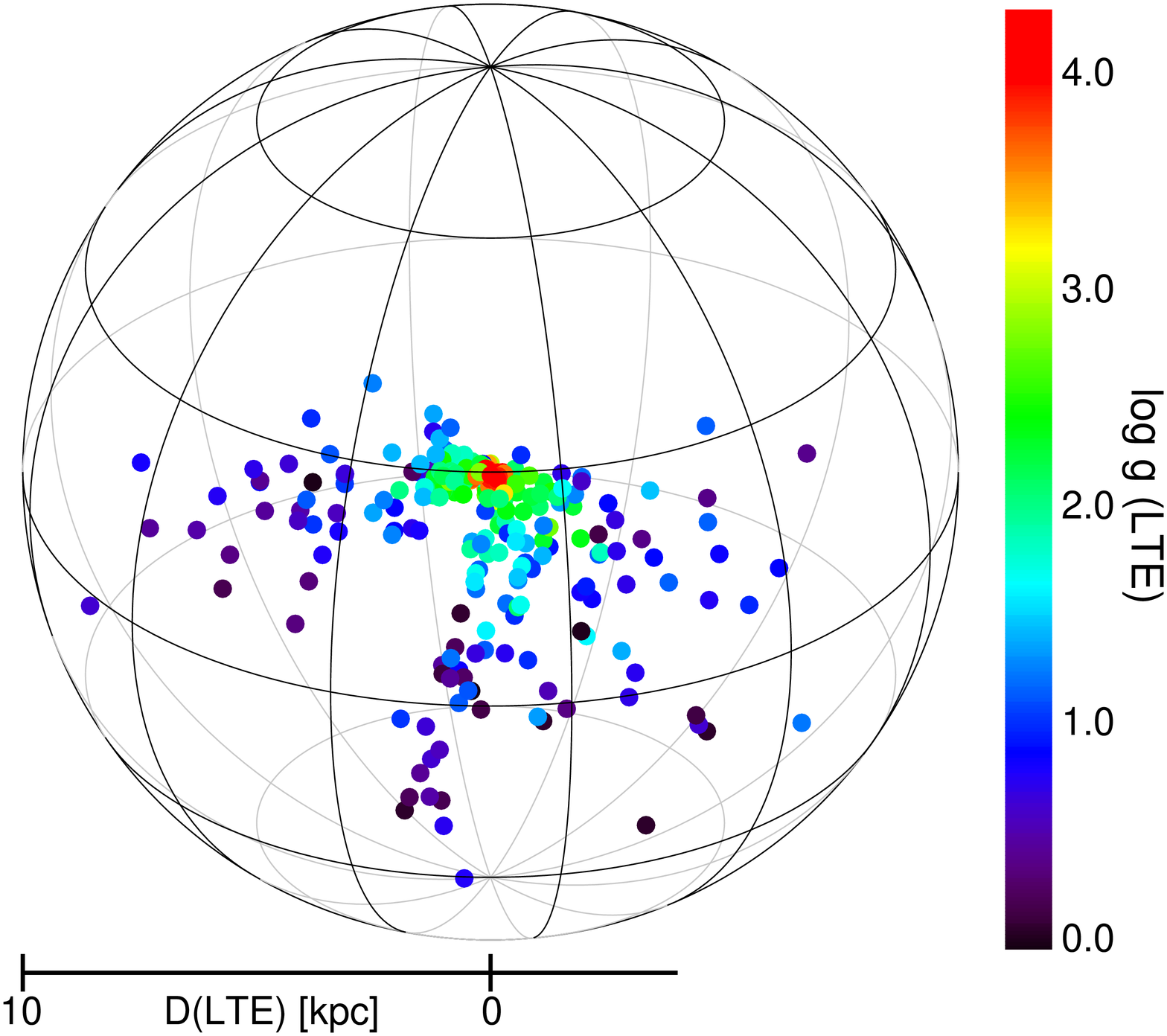} 
\includegraphics[width=90mm]{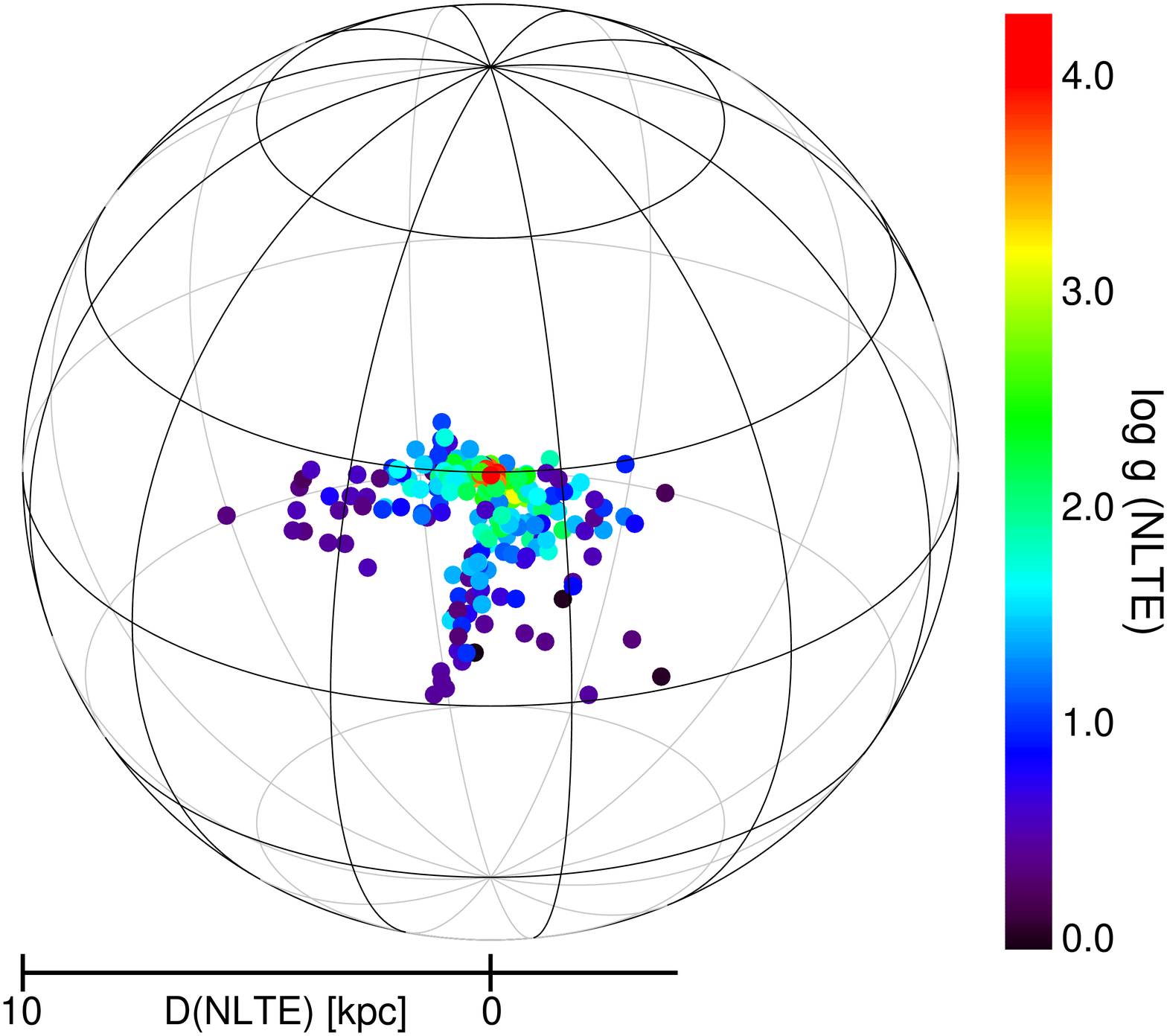}
\caption{Heliocentric spatial distribution of stars in our sample. Top
and bottom panels  respectively show  results  obtained based  on $\pone$  and
$\ptwo$ parameters. Colours depict the stellar surface gravity as indicated in
the plots. 
\label{fig:sphere}}
\end{figure}

In  Paper 1, we  developed a  novel NLTE-Opt  technique to  determine accurate
effective  temperature,  surface  gravity,  and  metallicity  from  a  stellar
spectrum and then applied the method  to a large sample of stars selected from
the RAVE survey. The  principal components of the method are (i)  the use of a
robust effective temperature scale from the analysis of Balmer line wings and,
(ii)  the determination  of  surface gravities  and  metallicities by  solving
physically-realistic  NLTE  radiative  transfer  in  Fe  lines.   This  avoids
substantial   systematic  uncertainties  in   parameters  introduced   by  the
commonly-used approximation of LTE.  A  comparison of the new NLTE-Opt stellar
parameters   to  that   obtained  from   the   standard  method   of  1D   LTE
excitation-ionization  balance of Fe,  LTE-Fe, revealed  important differences
ranging from  $-100$ K to $+400$  (subgiants, RGB) in $T_{\rm  eff}$, $0.1$ to
$1.5$ dex in $\logg$, and $0.05$ to $0.5$ dex in [Fe/H].

In this  paper, we  focus on the  determination of evolutionary  parameters of
stars,  masses and ages,  and their  distances. We  develop a  robust Bayesian
technique  using  state-of-the-art   evolutionary  tracks  computed  with  the
\texttt{GARSTEC}  code  \citep{weiss:2008}.  The  importance of  the  Bayesian
scheme  in   this  context  has   been  raised  in  several   earlier  studies
\citep{pont:2004,jorgensen:2005}. This  method has a  principal advantage over
standard maximum likelihood (or minimum  $\chi^2$) methods in that it allows a
simple  way of  introducing additional  (prior) knowledge  in  the statistical
analysis. Test comparisons  of results obtained with our  Bayesian scheme with
other similar studies confirm its reliability and accuracy.

We then apply  the Bayesian method to both sets  of stellar parameters, LTE-Fe
and  NLTE-Opt, for  the sample  of stars  from our  Paper 1.   The  stars were
selected  from the RAVE  catalogue.  They  cover the  full parameter  space in
$T_{\rm eff}$  ($3500$ to $6500$ K),  $\logg$ ($3500$ to  $6500$), and [Fe/H]
($-3.0$   to  $-0.5$)   and  sample   the  local   volume  within   $\sim  10$
kpc. Spectroscopic distances were computed for all these stars using the 2MASS
photometry. Masses  and ages were determined  for the $59$  dwarf and subgiant
stars ($\logg > 3.6$) only, the limit set by the degeneracy of stellar tracks
of different mass along the RGB in the $\Teff - \logg$ plane.

From these calculations, the following conclusions can be drawn.

\begin{figure}
\includegraphics[width=85mm]{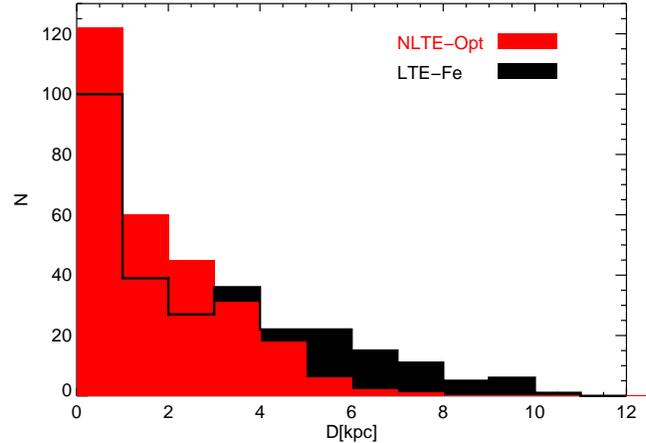}
\caption{Distribution  of distances  for our  full sample  showing the
    global impact of using the physically realistic $\ptwo$ stellar parameters
    to estimate distances.
\label{fig:dist_histo}}
\end{figure}

\begin{itemize}
 \item The  \textit{classical method} of LTE  excitation-ionization balance of
   Fe, as  a rule, underestimates ${\rm  Teff}$, $\logg$, and  [Fe/H] of FGKM
   stars. As demonstrated  in Paper 1, for the  most metal-poor giants, [Fe/H]
   $\sim -3$, the errors are of  the order $+0.5$ dex in metallicity. This has
   implications for the \textit{metallicity distribution function} of stars in
   different  Galactic   components.   For   our  stellar  sample,   which  is
   representative of  the thick-disk, the  metal-poor end of the  MDF steepens
   and is skewed  towards the more metal-rich domain  (by $\Delta$[Fe/H] $\sim
   +0.3$ dex; \citealp{bergemann:2013}).

 \item LTE-Fe stellar parameters result in overestimated distances compared to
   the  NLTE-Opt  results,  and  the  bias  depends  on  the  metallicity  and
   evolutionary  stage of a  star.  The  changes in  surface gravity  are most
   critical for  metal-poor RGB  stars, for which  the NLTE-Opt  distances are
   smaller by up to $70 \%$.  The distances to metal-poor subgiant and turnoff
   stars are in error by a  factor of $2$. Only more metal-rich turn-off stars
   are barely affected by  this bias.  Fig.~\ref{fig:sphere} shows the spatial
   heliocentric  distribution  of the  stars  in  our  sample determined  with
   $\pone$ (top panel) and $\ptwo$ (bottom panel) parameters and makes evident
   the  large  impact of  using  inaccurate  stellar  parameters for  deriving
   spectrophotometric    distances,   in    particular    for   giants    with
   $\logg<2.0$. The  overall distribution of distances in  our sample steepens
   towards  smaller distances  as  shown in  Fig.~\ref{fig:dist_histo}.  As  a
   consequence,  significant  errors  may  enter the  determination  of  space
   velocities and  orbital eccentricities,  and differences in  the population
   membership of stars (i.e. thick disk or halo) are expected.

\item   Comparing  our   NLTE-Opt  results   with  the   RAVE   DR3  distances
  \citep{zwitter:2010} for the stars common  in both samples we find a pattern
  similar to the  comparison between NLTE-Opt and LTE-Fe  results. This is not
  surprising: metal lines that dominate  the near-IR RAVE spectra are known to
  be strongly affected by NLTE effects (Sect.~\ref{sec:ravecomp}) and our test
  calculations with mean  3D models also show that the  lines are sensitive to
  the  ${\rm  1D-\left<3D\right>}$ atmospheric  structure  differences.  As  a
  result, more distant stars (predominantly metal-poor giants) have larger DR3
  distances  compared  to   our  values  (Fig.~\ref{fig:rave_vs_us_dist}),  by
  $10-50\%$.

  In this respect, it is  very interesting to interpret our results in
    the  light of  the findings  by \citet{schonrich:2011}  where  the authors
    concluded, based  on simple  analytical considerations that  distances for
    SDSS/SEGUE        metal-poor        stars        are        systematically
    overestimated.  Consistent with  the arguments  presented  in that
      study,  our calculations predict  that the  systematic error  in $\logg$
      and,  thus, distance  increases with  decreasing metallicity;  [Fe/H] is
      affected  too, although  at a  lesser degree.  Thus, there  will  be two
      effects.   First, more metal-poor  stars will  appear more  luminous and
      thus more  distant, which may  be mis-interpreted for a  very metal-poor
      outer stellar population.  This goes in the direction  of explaining why
      this component  would appear to be  more metal poor than  the inner halo
      component. Second,  the systematic error in distance  may propagate into
      the error  in transverse  velocity, causing  the effect of  a star  on a
      retrograde orbit. This  strengthens the claim by \citet{schonrich:2011}
    that the available observational  evidence for a counter-rotating Galactic
    halo component \citep{carollo:2010} is weak.

 \item  Masses and  ages are  likewise  affected in  a systematic  way by  the
   differences between NLTE-Opt and LTE-Fe, although there is not a one-to-one
   mapping of the change in mass to  the change in age.  In the LTE-Fe method,
   the  majority  of stars  appear  younger  and  slightly more  massive.  The
   difference with  NLTE-Opt is of the  order $0.05 - 0.1$  M$_{\rm sun}$. The
   ages are  inaccurate by  anything between $10$  and $30$ percent.   In this
   respect,    it    will     be    interesting    to    re-investigate    the
   \textit{age-metallicity}  relation  in  the  solar neighborhood,  which  is
   usually  drawn from  stellar samples  with spectroscopic  metallicities and
   ages obtained by simple isochrone fitting.

\end{itemize}

We emphasize that,  contrary to a wide-spread impression,  departures from LTE
and 1D  hydrostatic equilibrium will affect low-  and high-resolution spectra.
The reason is  that these effects change equivalent  widths (or, equivalently,
line  indices),  and  consequently   abundances  inferred  from  a  diagnostic
feature. This can  not be compensated by decreasing  resolution of a spectrum,
which merely alters the line  profile shape without affecting the total energy
absorbed in a line.

In  summary, we  have shown  that the  accuracy of  the stellar  parameters ($
\Teff$, $\logg$ and [Fe/H]) is crucial to obtain reliable masses, 
ages, and spectroscopic  distances for late-type stars.  Our  findings lead to
the  inevitable conclusion  that using  the adequate  $\Teff$  scale and
accounting  for the  deviations from  LTE  and 1D  hydrostatic equilibrium  in
stellar  atmospheres is of  fundamental importance  if we  want to  achieve an
accurate characterization  of stellar populations. Until  now, most of
  the work in the community has gone into developing sophisticated statistical
  methods  to  extract  the   largest  possible  amount  of  information  from
  large-scale  stellar  surveys.    Although  these  efforts  are  undoubtedly
  valuable,  they  will  not  cure  the  major  problem  affecting  the  final
  distributions  of  masses,  ages,  metallicities,  distances,  which  arises
  because  systematic errors in  basic stellar  parameters dominate.   We thus
  advise particular caution when  applying stellar parameters obtained from 1D
  LTE models in studies of stellar populations.

\section*{Acknowledgments}

We thank  the anonymous  referee for comments  that have helped  improving the
presentation of results.  AMS is partially supported by the European Union
International  Reintegration  Grant   PIRG-GA-2009-247732,  the  MICINN  grant
AYA2011-24704,  by  the  ESF   EUROCORES  Program  EuroGENESIS  (MICINN  grant
EUI2009-04170),  by SGR  grants of  the Generalitat  de Catalunya  and  by the
EU-FEDER  funds.  GRR  acknowledges partial  support through  grants  from ESF
EuroGenesis and Max Planck Society for the FirstStars collaboration.

\bibliographystyle{mn2e}

\label{lastpage}
\end{document}